\documentclass[prd,aps,nofootinbib,notitlepage,showpacs,showkeys,preprintnumbers]{revtex4-1}
\usepackage{graphicx,epsf,amsmath,amsfonts,amssymb,amsbsy}
\usepackage{epsfig}
\usepackage{epstopdf}

\newcommand{\ds}{\displaystyle}
\newcommand{\vev}[1]{\langle#1\rangle}
\newcommand{\mat}{\left ( \begin{array}}
\newcommand{\emat}{\end{array} \right )}
\newcommand{\vect}{\left ( \begin{array}{c}}
\newcommand{\evect}{\end{array} \right )}

\newcommand{\Det}{\mathop{\rm Det}\nolimits}

\begin{document}

\title{ \bf Dense baryon matter with isospin and chiral 
imbalance in the framework of NJL$_4$ model at large $N_c$: duality between chiral symmetry breaking and charged pion condensation }

\author{
T. G. Khunjua $^{1)}$, K. G. Klimenko $^{2)}$, and R. N. Zhokhov $^{2)}$
}
\vspace{1cm}

\affiliation{$^{1)}$ Faculty of Physics, Moscow State University,
119991, Moscow, Russia} \affiliation{$^{2)}$ Logunov Institute for High Energy Physics, NRC "Kurchatov Institute", \\142281, Protvino, Moscow Region, Russia}

\begin{abstract}
In this paper the phase structure of dense quark matter has been investigated at zero temperature in the presence of baryon, isospin and chiral isospin chemical potentials in the framework of massless (3+1)-dimensional Nambu--Jona-Lasinio model with two quark flavors. It has been shown that in the large-$N_{c}$ limit ($N_{c}$ is the number of colors of quarks) there exists a duality correspondence between the chiral symmetry breaking phase and the charged pion condensation one.
The key conclusion of our studies is the fact that chiral isospin chemical potential generates charged pion condensation in dense quark matter with isotopic asymmetry.
\end{abstract}


\maketitle

\section{Introduction}

In recent years great effort has been devoted to a problem of mapping out the phase diagram of QCD as a function of temperature and baryon chemical potential. However, theoretical investigations of QCD encounter considerable difficulties in the low-energy as well as low-temperature and density region, where perturbative methods do not work. The only possible first principle calculation in QCD at low energies is lattice approach to QCD. Unfortunately, its main method (Monte Carlo simulations) cannot be applied to investigations at finite baryon chemical potential due to the sign problem (complex fermion determinant). In order to study the phase diagram of QCD at nonzero chemical potential one usually use effective field theories such as chiral perturbation theory (ChPT), quark-meson model, etc. These theories are low energy approximations to QCD. In this case, the crucial point is to make an appropriate choice of degrees of freedom that are able to capture the physics which is most important for the problem at hand. For example, in ChPT one considers hadrons as elementary excitation of strongly interacting matter. And Lagrangian is obtained only from symmetry breaking pattern, etc. One of the most widely used effective theory is Nambu--Jona-Lasinio (NJL) model \cite{Nambu:1961fr} (see for review \cite{Klevansky:1992qe,Hatsuda:1994pi,Buballa:2003qv}). The degrees of freedom of this model are not hadrons as in ChPT but quarks. They are self-interacting and there are no gluons in considerations, they are integrated out. The model is tractable and can be used as low energy effective theory for QCD. The most attractive feature of NJL models is the dynamical breaking of the chiral symmetry (quarks acquire a relatively large mass) and hence it can be used as a basis model for constituent quark model.

Besides the temperature and baryon density (nonzero baryon chemical potential), there are additional parameters, which may be relevant for a description of the QCD phase diagram. These parameters are, in particular, an isotopic chemical potential and chiral chemical potential. Isotopic (isospin) chemical potential allows us to consider systems with isospin imbalance (different numbers of $u$ and $d$ quarks). In fact, nature provides us with systems at finite isospin chemical potential in the form of isospin asymmetric matter inside neutron stars. Moreover, in any heavy-ion experiment there is, even if small but nonzero, isospin asymmetry in the system of colliding ions. Also in the fireball after heavy-ion collisions there may emerge another very interesting phenomenon -- phase with a chiral imbalance of quarks, i.e., with different average numbers between right-handed and left-handed quarks. This phenomenon is a remarkable effect that stems from highly nontrivial interplay of chiral symmetry of QCD, axial anomaly, and the topology of gluon configurations, which may possibly
lead to the chiral magnetic effect \cite{fukus} in heavy-ion
collisions. It might be realized also in compact stars or condensed
matter systems \cite{andrianov} (see also the review \cite{ms}). Note
also that phenomena, connected with a chiral imbalance, are usually
described in the framework of NJL models with a chiral chemical
potential \cite{andrianov,braguta}.

It was contemplated a long time ago that at high densities there could be such a phenomenon as condensation of $\pi^0$ mesons \cite{Migdal,Tatsumi}. However, if there is an isospin imbalance, then the charged pion condensation (PC) phase might be realized in the system \cite{son}. Charged PC has been also considered in NJL-type models in \cite{eklim,ak,mu,andersen}. But the existence of the charged PC phase in dense baryon matter is predicted in the framework of NJL models without sufficient certainty. Indeed, for some values of model parameters (coupling constant $G$, cutoff parameter $\Lambda$, etc.) the charged PC phase with {\it nonzero baryon density} is allowed by NJL models. However, it is forbidden in the framework of NJL models for other physically interesting values of $G$ and $\Lambda$ \cite{eklim}. Moreover, if the electric charge neutrality constraint is imposed, the charged PC phenomenon depends strongly on the bare (current) quark mass values. In particular, it turns out that in this case the charged PC phase with
{\it nonzero baryonic density} is forbidden in the framework of NJL models, if the bare quark masses reach the physically acceptable values of $5\div 10$ MeV (see Refs \cite{andersen}). Due to these circumstances, the question arises whether there exist factors promoting the appearance of charged PC phenomenon in dense baryonic matter. A positive answer to this question was obtained in the papers \cite{ekkz,gkkz,ekk,kkzz}, where it was shown that a charged PC phase might be realized in dense baryonic system with finite size, or in the case of a spatially inhomogeneous pion condensate, or in the case of chiral imbalance in the system. These conclusions are demonstrated in Refs \cite{ekkz,gkkz,ekk,kkzz}, using a massless (1+1)-dimensional toy model with four-quark interactions.

In this paper we study the phase diagram of QCD, and the charged PC phenomenon in particular, in the framework of (3+1)-dimensional
NJL model with two quark flavors and in the presence of the baryon ($\mu_{B}$), isospin ($\mu_{I}$) as well as chiral isospin $\mu_{I5}$ chemical potentials, i.e. under heavy-ion experimental and/or compact star conditions. The consideration is performed in the chiral limit. This study is also inspired by exploration of the phase diagram of QCD in the framework of (1+1)-dimensional massless NJL model performed in Ref. \cite{ekk} (for the case of homogeneous ansatz for condensates) and Ref. \cite{kkzz} (in inhomogeneous case) at nonzero $\mu_{B}$, $\mu_{I}$ and $\mu_{I5}$. It has been shown in these papers that in (1+1)-dimensional NJL model chiral isospin chemical potential generates charged PC in the system, and there is a duality between charged pion condensation and chiral symmetry breaking (CSB) phenomena. \footnote{Note that
another kind of duality correspondence, the duality between CSB and
superconductivity, was demonstrated both in (1+1)- and
(2+1)-dimensional NJL models \cite{ekkz2}.}

Let us elaborate a little bit more on this duality. In the present consideration it means that if at $\mu_{I}=A$ and $\mu_{I5}=B$ (at arbitrary fixed chemical potential $\mu_{B}$), e.g., the CSB (or the charged PC) phase is realized in the model, then at the permuted values of these chemical potentials, i.e. at $\mu_{I}=B$ and $\mu_{I5}=A$, the charged PC (or the CSB) phase is arranged. So, it is enough to know the phase structure
of the model at $\mu_{I}<\mu_{I5}$, in order to establish the phase
structure at $\mu_{I}> \mu_{I5}$, and vice versa. Knowing condensates and other dynamical and thermodynamical quantities of the system, e.g., in the CSB phase, one can then obtain the corresponding quantities in the dually conjugated charged PC phase of the model, by simply performing there the duality transformation, $\mu_{I}\leftrightarrow\mu_{I5}$.

In our case duality is between condensates and matter content (chemical potentials). But generally, notion of duality is more widespread and is a very powerful tool and its use permeates all branches of physics. String theory, condensed matter physics, etc. If one finds duality between two theories and one theory is more thoroughly investigated than the other, duality can be used for applying known solutions of one theory to the other. This can saves the effort of the researchers. But dualities can be even more powerful in a case when one theory cannot be solved by existing methods (for example perturbation theory) and the other could be. For example, there are strong-weak dualities that connect weak coupling regime of one theory with strong coupling regime of the other. Such a duality is AdS/CFT (or gauge/gravity) duality \cite{Maldacena:1997re}, which relates some strongly-coupled, four-dimensional gauge theories at large $N_{c}$ to tractable weakly-coupled string theories living in ten dimensions. Now AdS/CFT conjecture is a subject of very intense study.

A different notion (but one historically connected to AdS/CFT) is that of strong-strong dualities, which usually go by the name of large-$N_{c}$ orbifold equivalences \cite{Cherman,1103.5480,Hanada:2011ju,Kashiwa:2017yvy}.
Orbifold equivalences connect gauge theories with different gauge groups and matter content in the large-$N_{c}$ limit. Let us elaborate on why it could be important. As we already said, the lattice technique is not available at finite $\mu_B$ because of the sign problem. Still, there is a class of QCD-like theories which are free from the sign problem. These theories intensively studied so far include two-color QCD with even numbers of fundamental flavors, any-color $SU(N_{c})$ QCD with adjoint fermions, and $SU(N_{c})$ QCD at finite isospin chemical potential $\mu_I$ \cite{han}. There are also other theories which are free from sign problem at finite $\mu_B$, for example, $SO(2N_{c})$ and $Sp(2N_{c})$ gauge theories with $N_f$ fundamental Dirac fermions. In Ref. \cite{1103.5480} it is argued that the whole or the part of the phase diagrams of QCD and $SO(2N_{c})$ and $Sp(2N_{c})$ gauge QCD-like theories should be universal in the large-$N_{c}$ limit via the orbifold equivalence. So if one can prove the equivalence of phase portraits of two theories one of which has no sign problem, one can explore the phase structure of the other one (which could have sign problem) using lattice QCD. In the framework of orbifold equivalence formalism in \cite{1103.5480} there has been also obtained a similar (to our) dualities between charged PC and chiral symmetry breaking phenomena. Namely, it was shown that the whole phase diagram of QCD at chiral chemical potential must be identical to that of QCD at isotopic chemical potential in the chiral limit, where the charged pion condensation is replaced by the chiral condensate. The result has been shown only for a large number of colors $N_{c}$, but it was argued that the universality may work approximately even for $N_{c}=3$.

In the present paper we show that in the chiral limit the main result of the (1+1)-dimensional NJL model consideration that chiral isospin chemical potential generates charged pion condensation holds in more realistic (3+1)-dimensional NJL model as well. Moreover, in (3+1)-dimensional NJL model with nonzero $\mu_{B}$, $\mu_{I}$ and $\mu_{I5}$ in the chiral limit there is also a duality correspondence between CSB and charged PC phenomena.

In our opinion, it is a very inspiring result that phase diagram of a toy  (1+1)-dimensional model for QCD and more realistic effective (3+1)-dimensional NJL theory for QCD looks very similar in the chiral limit and predicts the emergence of charged PC phase in dense isotopically and chirally asymmetric matter. This qualitative agreement makes one believe that the common features of this model phase diagram could be realized in real QCD as well.

The paper is organized as follows. First, in Sec. II a (3+1)-dimensional massless NJL model with two quark flavors ($u$ and $d$ quarks) that also includes three kinds of chemical potentials, $\mu_B,\mu_I,\mu_{I5}$, is introduced. Furthermore, the symmetries of the model are discussed and thermodynamic potential (TDP) of the model under consideration is presented in the leading order of the large-$N_c$ expansion. Here the duality (dual symmetry) of the model TDP is established. The essence of duality is that TDP is invariant under the simultaneous interchange of $\mu_I,\mu_{I5}$ chemical potentials as well as chiral and charged pion condensates.
In Sec. III it is argued that there is no mixed phase in the system (with nonzero chiral symmetry breaking $M$- and charged pion $\Delta$ condensates, simultaneously) and expressions for projections of TDP to the $M$- and $\Delta$ axes are presented. Sec. IV contains the discussion on the phase structure of the model under consideration and different phase portraits of the model are depicted. In the section, the fact that chiral isospin chemical potential generates charged pion condensation in dense quark matter with isotopic asymmetry is established. Moreover, here the role of duality between chiral symmetry breaking and charged pion condensation phenomena and its influence on the phase diagram are explained. In Sec V summary and conclusions are given. Some technical details are relegated to Appendix A.

\section{The model and its thermodynamic potential}

We study a phase structure of the two flavored (3+1)-dimensional massless NJL model with several chemical potentials. Its Lagrangian, which is symmetrical under global color $SU(N_c)$ group, has the form
\begin{eqnarray}
&&  L=\bar q\Big [\gamma^\nu\mathrm{i}\partial_\nu
+\frac{\mu_B}{3}\gamma^0+\frac{\mu_I}2 \tau_3\gamma^0+\frac{\mu_{I5}}2 \tau_3\gamma^0\gamma^5\Big ]q+ \frac
{G}{N_c}\Big [(\bar qq)^2+(\bar q\mathrm{i}\gamma^5\vec\tau q)^2 \Big
]  \label{1}
\end{eqnarray}
and describes dense baryonic matter with two massless $u$ and $d$ quarks, i.e. $q$ in (1) is the flavor doublet, $q=(q_u,q_d)^T$, where $q_u$ and $q_d$ are four-component Dirac spinors as well as color $N_c$-plets (the summation in (\ref{1}) over flavor, color, and spinor indices is implied); $\tau_k$ ($k=1,2,3$) are Pauli matrices. The Lagrangian (1) contains baryon $\mu_B$-, isospin $\mu_I$-, and chiral isospin $\mu_{I5}$ chemical potentials. In other words, this model is able to describe the properties of quark matter with nonzero baryon $n_B$-, isospin $n_I$-, and chiral isospin $n_{I5}$ densities which are the quantities, thermodynamically conjugated to chemical potentials $\mu_B$, $\mu_I$, and $\mu_{I5}$, respectively.

The quantities $n_B$, $n_I$ and $n_{I5}$ are densities of conserved charges, which correspond to the invariance of Lagrangian (1) with respect to the abelian $U_B(1)$, $U_{I_3}(1)$ and $U_{AI_3}(1)$ groups, where \footnote{\label{f1,1}
Recall for the following that~~
$\exp (\mathrm{i}\alpha\tau_3)=\cos\alpha
+\mathrm{i}\tau_3\sin\alpha$,~~~~
$\exp (\mathrm{i}\alpha\gamma^5\tau_3)=\cos\alpha
+\mathrm{i}\gamma^5\tau_3\sin\alpha$.}
\begin{eqnarray}
U_B(1):~q\to\exp (\mathrm{i}\alpha/3) q;~
U_{I_3}(1):~q\to\exp (\mathrm{i}\alpha\tau_3/2) q;~
U_{AI_3}(1):~q\to\exp (\mathrm{i}
\alpha\gamma^5\tau_3/2) q.
\label{2001}
\end{eqnarray}
So we have from (\ref{2001}) that $n_B=\bar q\gamma^0q/3$, $n_I=\bar q\gamma^0\tau^3 q/2$ and $n_{I5}=\bar q\gamma^0\gamma^5\tau^3 q/2$. We would like also to remark that, in addition to (\ref{2001}), Lagrangian (1) is invariant with respect to the electromagnetic $U_Q(1)$ group,
\begin{eqnarray}
U_Q(1):~q\to\exp (\mathrm{i}Q\alpha) q,
\label{2002}
\end{eqnarray}
where $Q={\rm diag}(2/3,-1/3)$. The ground state expectation values of $n_B$, $n_I$ and $n_{I5}$ can be found by differentiating the thermodynamic potential of the system (1) with respect to the corresponding chemical potential. The goal of the present paper is the investigation of the ground state properties (or phase structure) of the system (1) and its dependence on the chemical potentials $\mu_B$, $\mu_I$ and $\mu_{I5}$ (at zero temperature).

To find the thermodynamic potential of the system, we use a semibosonized version of the Lagrangian (\ref{1}), which contains
composite bosonic fields $\sigma (x)$ and $\pi_a (x)$ $(a=1,2,3)$
(in what follows, we use the notations
$\mu\equiv\mu_B/3$, $\nu=\mu_I/2$ and $\nu_{5}=\mu_{I5}/2$):
\begin{eqnarray}
\widetilde L\ds &=&\bar q\Big [\gamma^\rho\mathrm{i}\partial_\rho
+\mu\gamma^0
+ \nu\tau_3\gamma^0+\nu_{5}\tau_3\gamma^0\gamma^5-\sigma
-\mathrm{i}\gamma^5\pi_a\tau_a\Big ]q
 -\frac{N_c}{4G}\Big [\sigma\sigma+\pi_a\pi_a\Big ].
\label{2}
\end{eqnarray}
In (\ref{2}) and below the summation over repeated indices is implied. From the auxiliary Lagrangian (\ref{2}) one gets the equations
for the bosonic fields
\begin{eqnarray}
\sigma(x)=-2\frac G{N_c}(\bar qq);~~~\pi_a (x)=-2\frac G{N_c}(\bar q
\mathrm{i}\gamma^5\tau_a q).
\label{200}
\end{eqnarray}
Note that the composite bosonic field $\pi_3 (x)$ can be identified with the physical $\pi^0(x)$-meson field, whereas the physical $\pi^\pm (x)$-meson fields are the following combinations of the composite fields, $\pi^\pm (x)=(\pi_1 (x)\mp i\pi_2 (x))/\sqrt{2}$.
Obviously, the semibosonized Lagrangian $\widetilde L$ is equivalent to the initial Lagrangian (\ref{1}) when using the equations (\ref{200}).
Furthermore, it is clear from Eq. (\ref{2001}) and footnote \ref{f1,1} that the composite bosonic fields (\ref{200}) are transformed under the isospin $U_{I_3}(1)$ and axial isospin $U_{AI_3}(1)$ groups in the following manner:
\begin{eqnarray}
U_{I_3}(1):~&&\sigma\to\sigma;~~\pi_3\to\pi_3;~~\pi_1\to\cos
(\alpha)\pi_1+\sin (\alpha)\pi_2;~~\pi_2\to\cos (\alpha)\pi_2-\sin
(\alpha)\pi_1,\nonumber\\
U_{AI_3}(1):~&&\pi_1\to\pi_1;~~\pi_2\to\pi_2;~~\sigma\to\cos
(\alpha)\sigma+\sin (\alpha)\pi_3;~~\pi_3\to\cos
(\alpha)\pi_3-\sin (\alpha)\sigma.
\label{201}
\end{eqnarray}
Starting from the auxiliary Lagrangian (\ref{2}), one obtains in the leading order of the large-$N_c$ expansion (i.e. in the one-fermion loop
approximation) the following path integral expression for the
effective action ${\cal S}_{\rm {eff}}(\sigma,\pi_a)$ of the bosonic
$\sigma (x)$ and $\pi_a (x)$ fields:
$$
\exp(\mathrm{i}{\cal S}_{\rm {eff}}(\sigma,\pi_a))=
  N'\int[d\bar q][dq]\exp\Bigl(\mathrm{i}\int\widetilde L\,d^4 x\Bigr),
$$
where
\begin{equation}
{\cal S}_{\rm {eff}}
(\sigma(x),\pi_a(x))
=-N_c\int d^4x\left [\frac{\sigma^2+\pi^2_a}{4G}
\right ]+\tilde {\cal S}_{\rm {eff}},
\label{3}
\end{equation}
The quark contribution to the effective action, i.e. the term
$\tilde {\cal S}_{\rm {eff}}$ in Eq. (\ref{3}), is given by:
\begin{eqnarray}
\exp(\mathrm{i}\tilde {\cal S}_{\rm {eff}})&=&N'\int [d\bar
q][dq]\exp\Bigl(\mathrm{i}\int\Big\{\bar q\big
[\gamma^\rho\mathrm{i}\partial_\rho +\mu\gamma^0+
\nu\tau_3\gamma^0+\nu_5\tau_3\gamma^0\gamma^5-\sigma -\mathrm{i}\gamma^5\pi_a\tau_a\big
]q\Big\}d^4 x\Bigr)\nonumber\\
&=&[\Det D]^{N_c},
 \label{4}
\end{eqnarray}
where $N'$ is a normalization constant. Moreover, in Eq. (\ref{4}) we have introduced the notation $D$,
\begin{equation}
D\equiv\gamma^\nu\mathrm{i}\partial_\nu +\mu\gamma^0
+ \nu\tau_3\gamma^0+\nu_{5}\tau_3\gamma^0\gamma^5-\sigma (x) -\mathrm{i}\gamma^5\pi_a(x)\tau_a,
\label{5}
\end{equation}
for the Dirac operator, which acts in the flavor-, spinor- as well as coordinate spaces only. Using the general formula $\Det D=\exp {\rm Tr}\ln D$, one obtains for the effective action (\ref{3}) the following expression
\begin{equation}
{\cal S}_{\rm {eff}}(\sigma(x),\pi_a(x))
=-N_c\int
d^2x\left[\frac{\sigma^2(x)+\pi^2_a(x)}{4G}\right]-\mathrm{i}N_c{\rm
Tr}_{sfx}\ln D,
\label{6}
\end{equation}
where the Tr-operation stands for the trace in spinor- ($s$), flavor-
($f$) as well as four-dimensional coordinate- ($x$) spaces, respectively.

The ground state expectation values $\vev{\sigma(x)}$ and
$\vev{\pi_a(x)}$ of the composite bosonic fields are determined by
the saddle point equations,
\begin{eqnarray}
\frac{\delta {\cal S}_{\rm {eff}}}{\delta\sigma (x)}=0,~~~~~
\frac{\delta {\cal S}_{\rm {eff}}}{\delta\pi_a (x)}=0,
\label{05}
\end{eqnarray}
where $a=1,2,3$. Just the knowledge of $\vev{\sigma(x)}$ and
$\vev{\pi_a(x)}$ and, especially, of their behaviour vs chemical potentials supplies us with a phase structure of the model. It is clear from Eq. (\ref{201}) that if $\vev{\sigma(x)}\ne 0$ and/or $\vev{\pi_3(x)}\ne 0$, then the axial isospin $U_{AI_3}(1)$ symmetry of the model is spontaneously broken down, whereas at $\vev{\pi_1(x)}\ne 0$ and/or $\vev{\pi_2(x)}\ne 0$ we have a spontaneous breaking of the isospin $U_{I_3}(1)$ symmetry. Since in the last case the ground state expectation values, or condensates, both of the field $\pi^+(x)$ and of the field $\pi^-(x)$ are not zero, this phase is usually called the charged pion condensation (PC) phase. In addition, it is easy to see from Eq. (\ref{200}) that the nonzero condensates $\vev{\pi_{1,2}(x)}$ (or $\vev{\pi^\pm(x)}$) are not invariant with respect to the electromagnetic $U_Q(1)$ transformations (\ref{2002}) of the flavor quark doublet. Hence in the charged PC phase the electromagnetic $U_Q(1)$ invariance of the model (1) is also broken spontaneously, and superconductivity is an unavoidable property of the charged PC phase.

In the present paper we suppose that in the ground state of the system, i.e. in the state of thermodynamic equilibrium, the ground state expectation values $\vev{\sigma(x)}$ and $\vev{\pi_a(x)}$ do not depend on spacetime coordinates $x$,
\begin{eqnarray}
\vev{\sigma(x)}\equiv M,~~~\vev{\pi_a(x)}\equiv \pi_a, \label{8}
\end{eqnarray}
where $M$ and $\pi_a$ ($a=1,2,3$) are already constant quantities. In fact, they are coordinates of the global minimum point of the
thermodynamic potential (TDP) $\Omega (M,\pi_a)$.
In the leading order of the large-$N_c$ expansion it is defined, using Eq. (\ref{8}), by the following expression:
\begin{equation}
\int d^4x \Omega (M,\pi_a)=-\frac{1}{N_c}{\cal S}_{\rm
{eff}}\big (\sigma(x),\pi_a (x)\big )\Big|_{\sigma
(x)=M,\pi_a(x)=\pi_a} .\label{08}
\end{equation}
In what follows we are going to investigate the
$\mu,\nu,\nu_{5}$-dependence of the global minimum point of the function $\Omega (\sigma,\pi_a)$ vs $\sigma,\pi_a$. To simplify the task, let us note that due to a $U_{I_3}(1)\times U_{AI_3}(1)$ invariance of the model, the TDP (\ref{08}) depends effectively only on the two combinations, $\sigma^2+\pi_3^2$ and $\pi_1^2+\pi_2^2$, of the bosonic fields, as is easily seen from Eq. (\ref{201}). In this case, without loss of generality, one can put $\pi_2=\pi_3=0$ in Eq. (\ref{08}),
and study the TDP as a function of only two variables,
$M\equiv\sigma$ and $\Delta\equiv\pi_1$. So, throughout the paper we use the ansatz
\begin{eqnarray}
\vev{\sigma(x)}=M,~~~\vev{\pi_1(x)}=\Delta,~~~\vev{\pi_2(x)}=0,~~~ \vev{\pi_3(x)}=0. \label{06}
\end{eqnarray}
In this case the TDP (\ref{08}) reads
\begin{eqnarray}
\Omega (M,\Delta)~
&&=\frac{M^2+\Delta^2}{4G}+\mathrm{i}\frac{{\rm
Tr}_{sfx}\ln D}{\int d^4x}\nonumber\\
&&=\frac{M^2+\Delta^2}{4G}+\mathrm{i}\int\frac{d^4p}{(2\pi)^4}\ln\Det\overline{D}(p),
\label{07}
\end{eqnarray}
where
\begin{equation}
\overline{D}(p)=\not\!p +\mu\gamma^0
+ \nu\tau_3\gamma^0+ \nu_{5}\tau_3\gamma^0\gamma^5-M
-\mathrm{i}\gamma^5\Delta\tau_1\equiv\left
(\begin{array}{cc}
A~, & U\\
V~, & B
\end{array}\right )
\label{500}
\end{equation}
is the momentum space representation of the Dirac operator $D$ (\ref{5}) under the constraint (\ref{06}). The quantities $A,B,U,V$ in Eq. (\ref{500}) are really the following 4$\times$4 matrices,
\begin{equation}
A=\not\!p +\mu\gamma^0
+ \nu\gamma^0+ \nu_{5}\gamma^0\gamma^5-M;~~B=\not\!p +\mu\gamma^0
- \nu\gamma^0- \nu_{5}\gamma^0\gamma^5-M;~~U=V=-\mathrm{i}\gamma^5\Delta,
\label{80}
\end{equation}
so the quantity $\overline{D}(p)$ from Eq. (\ref{500}) is indeed a 8$\times$8 matrix whose determinant appears in the expression (\ref{07}). Based on the following general relations
\begin{eqnarray}
\Det\overline{D}(p)\equiv\det\left
(\begin{array}{cc}
A~, & U\\
V~, & B
\end{array}\right )=\det [-VU+VAV^{-1}B]=\det
[BA-BUB^{-1}V]
\label{9}
\end{eqnarray}
and using any program of analytical calculations, one can find from Eqs (\ref{80}) and (\ref{9})
\begin{eqnarray}
\Det\overline{D}(p)=\big (\eta^4-2a\eta^2-b\eta+c\big )\big (\eta^4-2a\eta^2+b\eta+c\big )\equiv P_-(p_0)P_+(p_0),
\label{91}
\end{eqnarray}
where $\eta=p_0+\mu$, $|\vec p|=\sqrt{p_1^2+p_2^2+p_3^2}$ and
\begin{eqnarray}
a&&=M^2+\Delta^2+|\vec p|^2+\nu^2+\nu_{5}^2;~~b=8|\vec p|\nu\nu_{5};\nonumber\\
c&&=a^2-4|\vec p|^2(\nu^2+\nu_5^2)-4M^2\nu^2-4\Delta^2\nu_5^2-4\nu^2\nu_5^2.
\label{10}
\end{eqnarray}
It is evident from Eq. (\ref{10}) that the TDP (\ref{07}) is an even
function over each of the variables $M$ and $\Delta$, and parameters $\nu$ and $\nu_5$. In addition, it is invariant under the transformation $\mu\to-\mu$. \footnote{Indeed, if simultaneously with $\mu\to-\mu$ we perform in the integral (\ref{07}) the $p_0\to-p_0$ change of variables, then one can easily see that the expression (\ref{07}) remains intact. }
Hence, without loss of generality we can consider in the following only $\mu\ge 0$, $\nu\ge 0$, $\nu_5\ge 0$, $M\ge 0$, and $\Delta\ge 0$ values of these quantities. Moreover, the TDP (\ref{07}) is invariant with respect to the so-called duality transformation \footnote{Let us note that the duality is present only in the chiral limit (zero current quark masses), in the real situation, when chiral symmetry is only approximate due to non-zero quark masses in the Lagrangian, duality will be only approximate as well. We will discuss it in one of the future works. }, 
\begin{eqnarray}
{\cal D}:~~~~M\longleftrightarrow \Delta,~~\nu\longleftrightarrow\nu_5.
 \label{16}
\end{eqnarray}
One can find roots of the polynomials (\ref{91}) analytically, the procedure is relegated to Appendix A. Four roots of $P_{+}(\eta)$ have the following form
\begin{eqnarray}
\eta_{1}=\frac{1}{2} \left(-\sqrt{r^2-4 q}-r\right),&&
~~~\eta_{2}=\frac{1}{2} \left(\sqrt{r^2-4 q}-r\right),\nonumber\\
\eta_{3}=\frac{1}{2} \left(r-\sqrt{r^2-4 s}\right),&&
~~~\eta_{4}=\frac{1}{2} \left(r+\sqrt{r^2-4 s}\right). \label{01}
\end{eqnarray}
The roots of $P_{-}(\eta)$ can be otained by changing $b\to-b$ (changing $b\to-b$ is equivalent to $q\leftrightarrow s$),
\begin{eqnarray}
\eta_{5}=\frac{1}{2} \left(-\sqrt{r^2-4 s}-r\right)=-\eta_{4},&&~~~
\eta_{6}=\frac{1}{2} \left(\sqrt{r^2-4 s}-r\right)=-\eta_{3},\nonumber\\
\eta_{7}=\frac{1}{2} \left(r-\sqrt{r^2-4 q}\right)=-\eta_{2},&&~~~
\eta_{8}=\frac{1}{2} \left(r+\sqrt{r^2-4 q}\right)=-\eta_{1}.\label{02}
\end{eqnarray}
where $q=\frac{1}{2} \left(-2 a+r^2-\frac{b}{r}\right)
,\,\,\,s=\frac{1}{2} \left(-2 a+r^2+\frac{b}{r}\right)$, and $r$ has quite complicated form, but could be always chosen as a real one (all the details can be found in Appendix A). So, it is evident from Eqs (\ref{07}) and (\ref{91}) that for the TDP one can obtain the following expression
\begin{eqnarray}
\Omega (M,\Delta)~
=\frac{M^2+\Delta^2}{4G}+\mathrm{i}\sum_{i=1}^{8}\int\frac{d^4p}{(2\pi)^4}\ln(p_{0}+\mu-\eta_{i}).\label{070}
\end{eqnarray}
Then, taking in account a general formula
\begin{eqnarray}
\int_{-\infty}^\infty dp_0\ln\big
(p_0-K)=\mathrm{i}\pi|K|,\label{int}
\end{eqnarray}
and using the fact that each root $\eta_i$ of Eqs (\ref{01}) and (\ref{02}) has a counterpart with opposite sign and a relation $|\mu-\eta_{i}|+|\mu+\eta_{i}|=2|\eta_{i}|+2\theta(\mu-|\eta_{i}|)(\mu-|\eta_{i}|) $, one gets
\begin{eqnarray}
\Omega (M,\Delta)
&=&\frac{M^2+\Delta^2}{4G}-\sum_{i=1}^{4}\int\frac{d^3p}{(2\pi)^3}\big (|\eta_{i}|+\theta(\mu-|\eta_{i}|)(\mu-|\eta_{i}|)\big )\nonumber\\
&=&\frac{M^2+\Delta^2}{4G}-\frac{1}{2\pi^2}\sum_{i=1}^{4}\int_{0}^{\Lambda}p^2\big (|\eta_{i}|+\theta(\mu-|\eta_{i}|)(\mu-|\eta_{i}|)\big )dp.\label{26}
\end{eqnarray}
To obtain the second line of Eq. (\ref{26}), where $p\equiv|\vec p|$ and $\Lambda$ is a three-momentum cutoff parameter, we have integrated in the first line of it over angle variables. In the following we will study the behaviour of the global minimum point of the TDP (\ref{26}) vs chemical potentials $\mu$, $\nu$ and $\nu_5$ for a special set of the model parameters,
$$
G=15.03\, GeV^{-1},\,\,\,\,\,\,\,\,\,\,\,\,\,\,\,\,\Lambda=0.65\, GeV.
$$
In this case at $\mu=0$, $\nu=0$ and $\nu_5=0$ one gets for constituent quark mass the value $M=301.58\, MeV$.
The same parameter set has been used, e.g., in Refs \cite{Buballa:2003qv,eklim}.

\section{Calculation of the TDP}
\subsection{Projections onto $M$ and $\Delta$ axes}

In the particular case when $\nu_5=0$ the roots $\eta_i$ (\ref{01}) of the polynomial $P_+(\eta)$ can be obtained in an explicit form (see, e.g., in Ref. \cite{eklim}), which is significantly simplifies the investigation of the TDP . Note that in the most general case when $\mu\ge 0$, $\nu\ge 0$, $\nu_5\ge 0$, $M\ge 0$, and $\Delta\ge 0$ the roots $\eta_i$ can be found analytically as well (see the procedure presented in Appendix A). However, the exact expressions for the roots $\eta_i$ and hence for the TDP have a rather cumbersome form, so we have not even show them in the paper, but the behavior of the TDP can be studied using numerical simulations. By this way it is possible to show that the TDP (\ref{26}) (as a function of $M$ and $\Delta$) can never has a global minimum point (GMP) of the form $(M\ne 0,\Delta\ne 0)$. It means that at arbitrary fixed values of chemical potentials $\mu\ge 0$, $\nu\ge 0$ and $\nu_5\ge 0$ the phase with nonzero both chiral and charged pion condensates cannot be realized in the model. So the GMP of the TDP (\ref{26}) lies either on the $M$ axis or on the $\Delta$ axis. Hence, in order to establish the phase portrait of the model, it is enough to study the projections $F_1(M)\equiv\Omega (M,\Delta=0)$ and $F_2(\Delta)\equiv\Omega(M=0,\Delta)$ of the TDP (\ref{26}) on the $M$ and $\Delta$ axes, correspondingly.

The roots of the polynomial $P_{+}(\eta)$ for the case of $M\ge 0$ and $\Delta=0$ have the following form
\begin{eqnarray}
\eta_{1,2M}\equiv\eta_{1,2}\Big |_{\Delta =0}=-\nu \pm\sqrt{M^2+(|\vec p|- \nu _5)^{2} },\,\,\,\,\,\,\,\,\,\,\,\,\,\,\,\,\,
\eta_{3,4M}\equiv\eta_{3,4}\Big |_{\Delta =0}=\nu \pm\sqrt{M^2+(|\vec p|+ \nu _5)^{2} }.
\label{etaCSB}
\end{eqnarray}
The expressions (\ref{etaCSB}) were obtained earlier in the paper \cite{ekk}, where we have studied a (1+1)-dimensional variant of the NJL model (1) also at $\mu\ne 0$, $\nu\ne 0$ and $\nu_5\ne 0$. Substituting Eqs (\ref{etaCSB}) into Eq. (\ref{26}), one can find the projection $F_1(M)$ of the TDP (\ref{26}) on the $M$ axis,
\begin{eqnarray}
F_{1}(M)=\Omega(M,0)
=\frac{M^2}{4G}-\frac{1}{2\pi^2}\sum_{i=1}^{4}\int_{0}^{\Lambda}p^2\big (|\eta_{iM}|+\theta(\mu-|\eta_{iM}|)(\mu-|\eta_{iM}|)\big )dp.
\label{F1ref}
\end{eqnarray}
The roots of $P_{+}(\eta)$ for the case of $M=0$ and $\Delta\ne 0$ have the following form (for details, see also the paper \cite{ekk})
\begin{eqnarray}
\eta_{1,2\Delta}\equiv\eta_{1,2}\Big |_{M =0}=-\nu_{5}\pm \sqrt{\Delta^2+(|\vec p|- \nu )^{2} },\,\,\,\,\,\,\,\,\,\,\,\,\,\,\,\,\,
\eta_{3,4\Delta}\equiv\eta_{3,4}\Big |_{M =0}=\nu_{5} \pm\sqrt{\Delta^2+(|\vec p|+ \nu )^{2} }.
\label{etaPC}
\end{eqnarray}
(Note that each root $\eta_{iM}$ of Eq. (\ref{etaCSB}) is conjugated to corresponding root $\eta_{i\Delta}$ of Eq. (\ref{etaPC}) with respect to the duality transformation (\ref{16}).) Then, the projection of thermodynamic potential to the axis $\Delta$ has the form
\begin{eqnarray}
F_{2}(\Delta)=\Omega(0,\Delta)
=\frac{\Delta^2}{4G}-\frac{1}{2\pi^2}\sum_{i=1}^{4}\int_{0}^{\Lambda}p^2\big (|\eta_{i\Delta}|+\theta(\mu-|\eta_{i\Delta}|)(\mu-|\eta_{i\Delta}|)\big )dp.
\label{F2ref}
\end{eqnarray}
The integration in Eqs (\ref{F1ref}) and (\ref{F2ref}) can be carried out analytically but the obtained expressions would be rather involved. So it is still easier to use numerical calculations for evaluation of the integrals.

As a result, we see that in order to find the GMP of the whole TDP (\ref{26}), one should compare the least values of the functions $F_1(M)$ and $F_2(\Delta)$. By this way, it is clear that there can exist no more than three different phases in the model (1). The first one is the symmetric phase, which corresponds to the global minimum point $(M_0,\Delta_0)$ of the TDP (\ref{26}) of the form $(M_0=0,\Delta_0=0)$. In the CSB phase the TDP
reaches the least value at the point $(M_0\ne 0,\Delta_0=0)$. Finally,
in the charged PC phase the global minimum point lies at the point $(M_0=0,\Delta_0\ne 0)$. (Notice, that in the most general case the coordinates (condensates) $M_0$ and $\Delta_0$ of the global minimum point depend on chemical potentials.)

\subsection{Quark number density}

Since the main goal of the present paper is to prove the possibility of the charged PC phenomenon in dense quark matter (at least in the framework of the NJL model (1)), the consideration of the physical quantity $n_{q}$, called quark number density, is now in order.
This quantity is a very important characteristic of the ground state. It is related to the baryon number density as $n_{q}=3n_B$ because $\mu=\mu_B/3$. Let us present here the ways how expressions for n$_{q}$ can be found in different phases.
Recall that in the general case this quantity is defined by the relation
\begin{eqnarray}
n_q=-\frac{\partial\Omega(M_0,\Delta_0)}{\partial\mu}, \label{37}
\end{eqnarray}
where $M_0$ and $\Delta_0$ are coordinates of the GMP of a thermodynamic potential. So in the chiral symmetry breaking phase we have
\begin{align}
n_q(\mu,\nu,\nu_{5})\bigg |_{CSB}=-\frac{\partial\Omega
(M_0,\Delta_0=0)}{\partial\mu}=-\frac{\partial
F_1(M_0)}{\partial\mu}.
\label{38}
\end{align}
Taking into account (\ref{F1ref}) it is not very
difficult to get the following expression
\begin{eqnarray}
n_{q}(\mu,\nu,\nu_{5})\bigg |_{CSB}
=\frac{1}{2\pi^2}\sum_{i=1}^{4}\int_{0}^{\Lambda}dpp^2\theta(\mu-|\eta_{iM_0}|),
\label{33}
\end{eqnarray}
where $\eta_{iM_0}$ is given by Eq. (\ref{etaCSB}) at $M=M_0$.

In a similar way, the particle density in the charged pion condensation phase looks like
\begin{align}
n_q(\mu,\nu,\nu_{5})\bigg |_{PC}=-\frac{\partial\Omega
(M_0=0,\Delta_0)}{\partial\mu}=-\frac{\partial
F_2(\Delta_0)}{\partial\mu}.
\end{align}
Since the quantity $F_2(\Delta_0)$ is defined by Eq. (\ref{F2ref}) at $\Delta=\Delta_0$, one can get
\begin{eqnarray}
n_{q}(\mu,\nu,\nu_{5})\bigg |_{PC}
=\frac{1}{2\pi^2}\sum_{i=1}^{4}\int_{0}^{\Lambda}dpp^2\theta(\mu-|\eta_{i\Delta_0}|),
\label{35}
\end{eqnarray}
where $\eta_{i\Delta_0}$ is defined by Eq. (\ref{etaPC}) at $\Delta=\Delta_0$.

\section{Phase diagram}

\subsection{Duality between chiral symmetry breaking and charged pion condensation}

In order to get phase structure of the model one has to find GMP $(M_0,\Delta_0)$ of the thermodynamic potential (\ref{26}). It has been already said earlier that there is no mixed phase, which corresponds to both $M_0\ne 0$ and $\Delta_0\ne 0$, and one can use projections $F_1(M)$ (\ref{F1ref}) and $F_2(\Delta)$ (\ref{F2ref}) of this TDP to the axes $M$ and $\Delta$, respectively. So it is necessary to determine the GMPs of these projections with respect to $M$ and $\Delta$. Then, one should compare the minimum values of these functions, the result is the GMP of the whole TDP (\ref{26}). In a physical sense, one should determine which of the minima is a real vacuum of the system and which is a metastable state. After this, using numerical calculations, it is necessary to study the behavior of the TDP global minimum point $(M_0,\Delta_0)$ vs chemical potentials. The result is the most general $(\mu,\nu,\nu_{5})$-phase portrait of the model, i.e. the one-to-one correspondence between any point $(\nu,\nu_5,\mu)$ of the three-dimensional space of chemical potentials and possible model phases (CSB, charged PC and symmetric phase). However, in order to obtain a more deep understanding of the phase diagram as well as to get a greater visibility of it, it is very convenient to consider different cross-sections of this general $(\mu,\nu,\nu_{5})$-phase portrait by the planes of the form $\nu= const$, $\nu_5= const$ and $\mu= const$.

In the next subsections these different cross-sections of the most general phase portrait will be presented. In addition, we will compare our results with corresponding phase diagrams of NJL$_2$ model. But before that, let us discuss the role and influence of the duality invariance (\ref{16}) of the model on the phase structure and demonstrate the ways how one can use it to obtain some phase portraits from the others without having to calculate anything.

Suppose that at some fixed particular values of chemical potentials $\mu$, $\nu=A$ and $\nu_5=B$ the global minimum of the TDP (\ref{26}) lies at the
point, e.g., $(M=M_0\ne 0,\Delta=0)$ . It means that for such fixed values of the chemical potentials the chiral symmetry breaking (CSB) phase is
realized in the model. Then it follows from the invariance of the TDP
with respect to the duality transformation ${\cal D}$ (\ref{16}) that at permuted chemical potential values (i.e. at $\nu=B$ and $\nu_5=A$ and intact value of $\mu$) the global minimum of the TDP $\Omega(M,\Delta)$ is arranged at the point $(M=0,\Delta=M_0)$, which corresponds to the charged PC phase (and vice versa). This is the so-called duality correspondence between CSB and charged PC phases in the framework of the model under consideration. Hence, the knowledge of a phase of the model (1) at some fixed values of external free model parameters $\mu,\nu,\nu_5$ is sufficient to understand what a phase (we call it a dually conjugated) is realized at rearranged values of external parameters, $\nu\leftrightarrow\nu_5$, at fixed $\mu$. Moreover, different physical parameters such as condensates, densities, etc, which characterize both the initial phase and the dually conjugated one, are connected by the duality transformation ${\cal D}$. For example, the chiral condensate of the initial CSB phase at some fixed $\mu,\nu,\nu_5$ is equal to the charged-pion condensate of the dually conjugated charged PC phase, in which one should perform the replacement $\nu\leftrightarrow\nu_5$. Knowing the particle density $n_q(\nu,\nu_{5})$ of the initial CSB phase as a function of chemical potentials $\nu,\nu_{5}$, one can find the particle density in the dually conjugated charged PC phase by interchanging $\nu$ and $\nu_{5}$ in the expression $n_q(\nu,\nu_{5})$, etc.

The duality transformation ${\cal D}$ of the TDP can also be applied to an arbitrary phase portrait of the model. In particular, it is clear that if we have a most general $(\nu,\nu_5,\mu)$-phase portrait, then under the duality transformation (which is now understood as a renaming both of the diagram axes, i.e. $\nu\leftrightarrow\nu_5$, and phases, i.e. CSB$\leftrightarrow$charged PC) this phase portrait is mapped to itself, i.e. the most general $(\nu,\nu_5,\mu)$-phase portrait is self-dual. Furthermore, the self-duality of the $(\nu,\nu_5,\mu)$-phase portrait means that in the three-dimensional $(\nu,\nu_5,\mu)$ space the regions of the CSB and charged PC phases are arranged mirror-symmetrically with respect to the plane $\nu=\nu_5$ of this space.

\subsection{$(\nu,\nu_{5})$-phase diagrams}

First, let us consider $\mu= const$ cross-sections of the general $(\nu,\nu_5,\mu)$-phase diagram. The result is different $(\nu,\nu_{5})$-phase diagrams at several typical values of the chemical potential $\mu$. They are depicted in Figs 1-8. We hope that from these plots it is possible to comprehend the behavior of the general phase diagram at all possible values of the quark number chemical potential $\mu$.

At zero $\mu$ the phase portrait is depicted in Fig. 1. It has charged PC phase which is arranged alone the $\nu$ axis at not very large $\nu_{5}$ values. Moreover, it is clear that there is also the CSB phase (at rather small values of $\nu$), which is arranged mirror-symmetrically to the charged PC phase with respect to the line $\nu=\nu_5$. Note that at $\mu=0$ baryon density $n_B\equiv n_q/3$ is zero in both phases. Let us recall that NJL model is not renormalizable and hence is an effective theory, which depends on a cutoff parameter $\Lambda$ (recall that in our case  $\Lambda=0.65$ GeV). Therefore, transitions to the symmetric phase at rather large $\nu$ and $\nu_{5}$ in Fig. 1, and also on other similar drawings, can be considered as an artifact of the cutoff theory. Clearly, this phase portrait supports the result of the papers \cite{braguta} obtained in lattice simulations that chiral imbalance generates chiral symmetry breaking in the system. \footnote{Strictly speaking, in Refs \cite{braguta} different systems with nonzero chiral $\mu_5$ chemical potential were studied, whereas we deal with chiral {\it isospin} chemical potential $\mu_{5I}$ (see Eq. (1)).}

When $\mu$ changes in the interval $0<\mu\lesssim 0.3$ GeV, then some hollow  appears in charged PC phase (see in Figs 2 and 3). Above the hollow one can see the charged PC$_{d}$ phase with nonzero baryon density (subscript $d$ means that baryon density is nonzero in the phase), whereas at the values of $\nu_{5}$ corresponding to a hollow there is still charged PC phase with zero baryon density. It means that at rather small values of $\mu$ the model predicts the charged PC phenomenon both in medium with $n_{B}=0$ (it might, e.g., consists of charged pions, etc) and in dense quark matter with $n_{B}\ne 0$, depending on the $\nu_5$ values. According to the duality symmetry (\ref{16}) of the model, in these figures the CSB phases (with $n_B=0$ and $n_B\ne 0$) are arranged on the other side (mirror-symmetrically) of a straight line $\nu=\nu_5$ as well.

\begin{figure}
\includegraphics[width=0.45\textwidth]{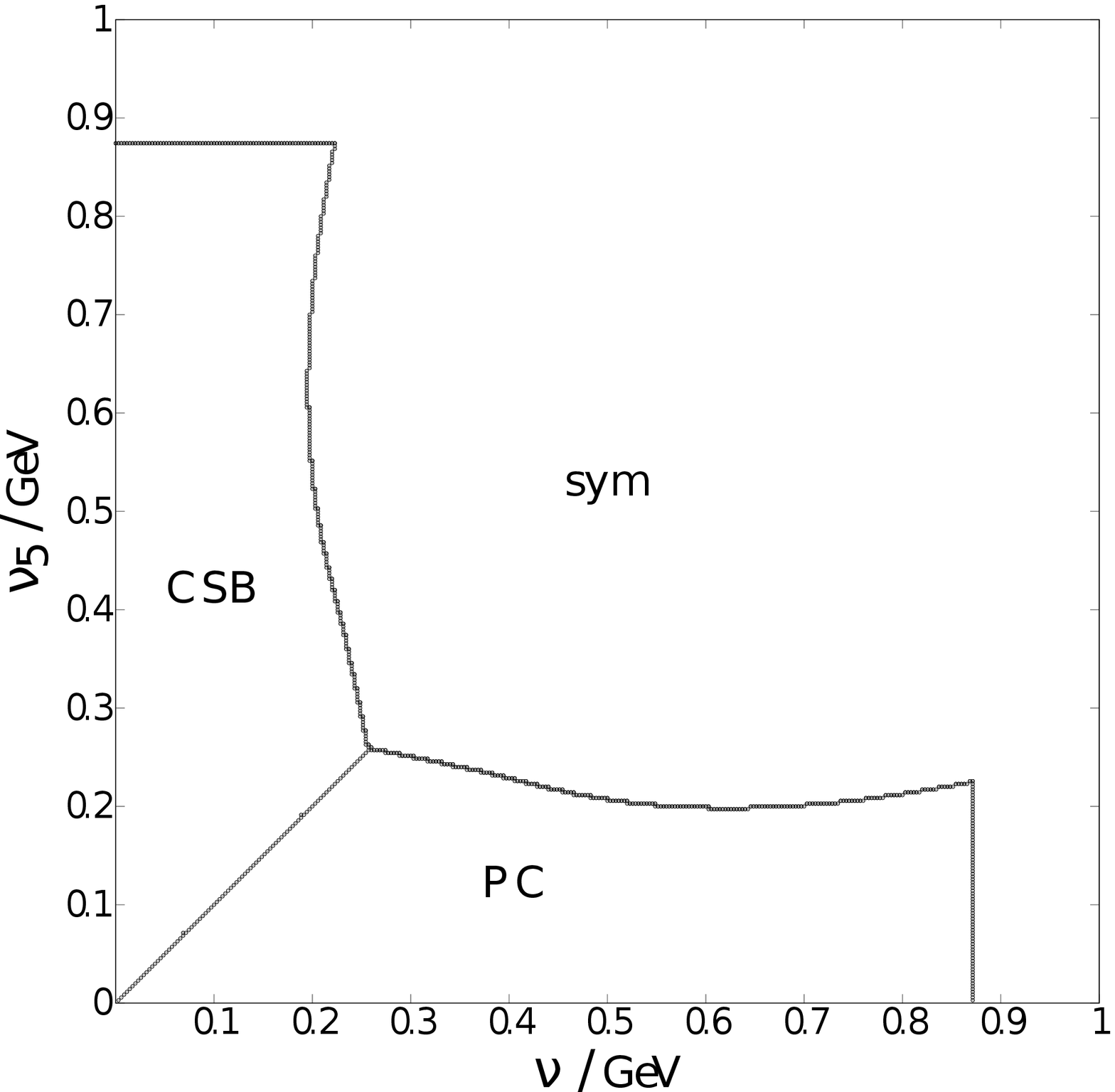}
 \hfill
\includegraphics[width=0.45\textwidth]{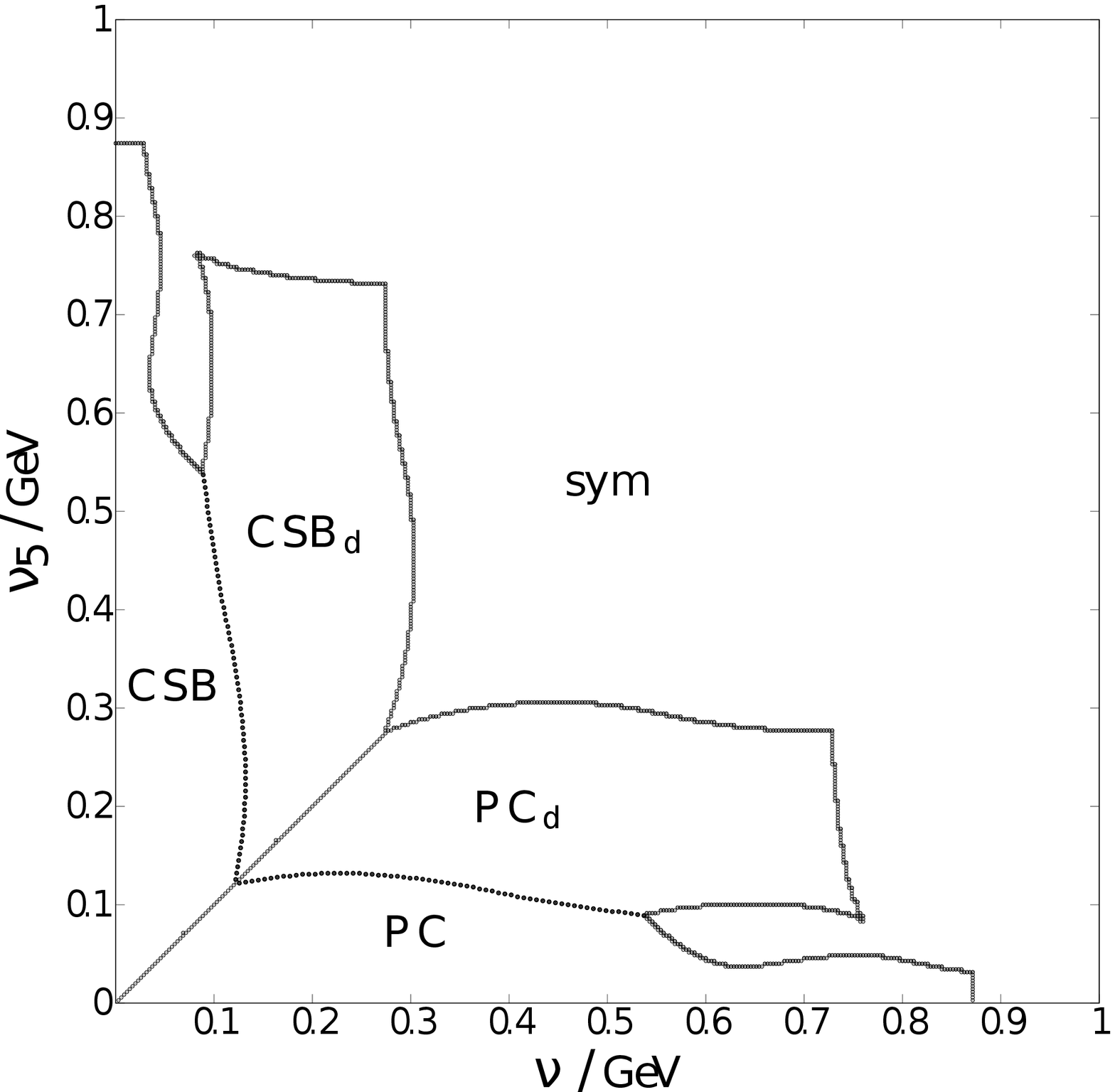}\\
\label{fig1}
\parbox[t]{0.45\textwidth}{
\caption{ $(\nu,\nu_{5})$-phase diagram at $\mu=0$ GeV. The notations CSB and PC mean, respectively, the chiral symmetry breaking and charged pion condensation phase with zero baryon density. The notation 'sym' stands for the symmetric phase, where all symmetries are restored.}
 }\hfill
\parbox[t]{0.45\textwidth}{
\caption{ $(\nu,\nu_{5})$-phase diagram at $\mu=0.195$ GeV. In addition to the notations of Fig. 1, here PC$_{d}$ and CSB$_{d}$ mean, respectively, the charged pion condensation and chiral symmetry breaking phase with nonzero baryon density.} }
\label{fig2}
\end{figure}
\begin{figure}
\includegraphics[width=0.45\textwidth]{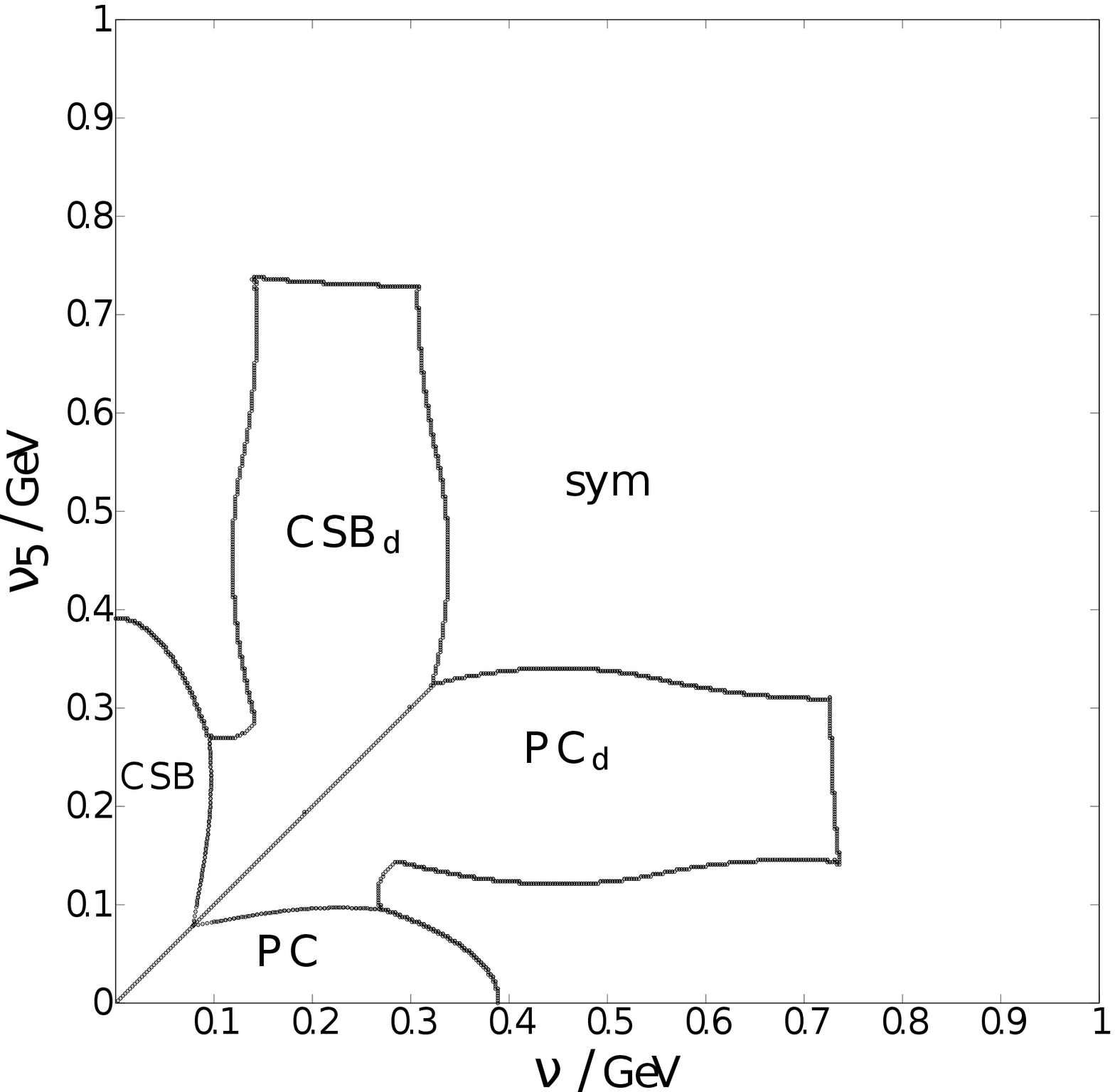}
 \hfill
\includegraphics[width=0.45\textwidth]{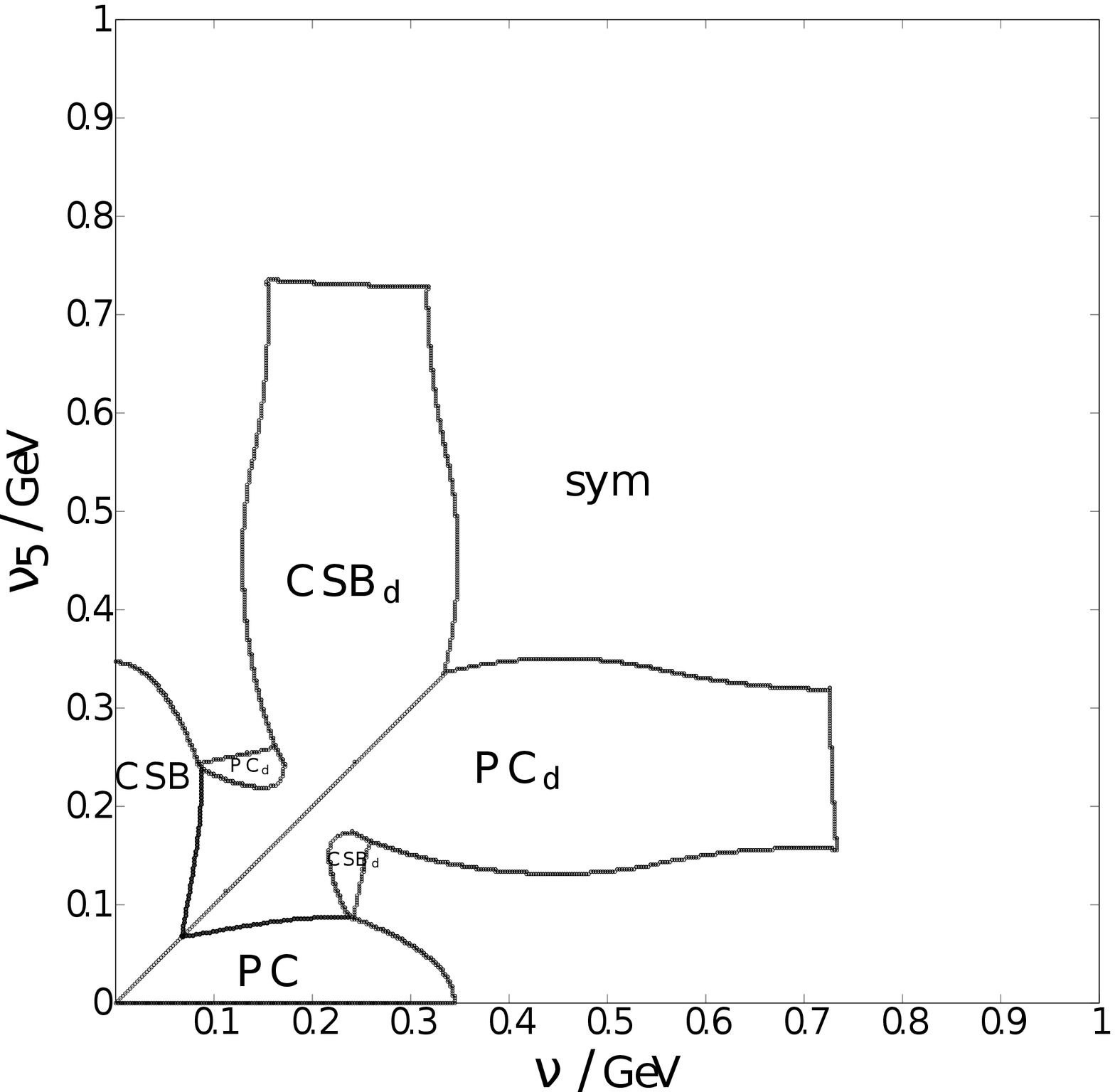}\\
\label{fig3}
\parbox[t]{0.45\textwidth}{
 \caption{ $(\nu,\nu_{5})$-phase diagram at $\mu=0.23$ GeV.  All the notations are the same as in Fig. 2.}
 }\hfill
\parbox[t]{0.45\textwidth}{
\caption{ $(\nu,\nu_{5})$-phase diagram at $\mu=0.24$ GeV. All the notations are the same as in Fig. 2.} }
\label{fig4}
\end{figure}
\begin{figure}
\includegraphics[width=0.45\textwidth]{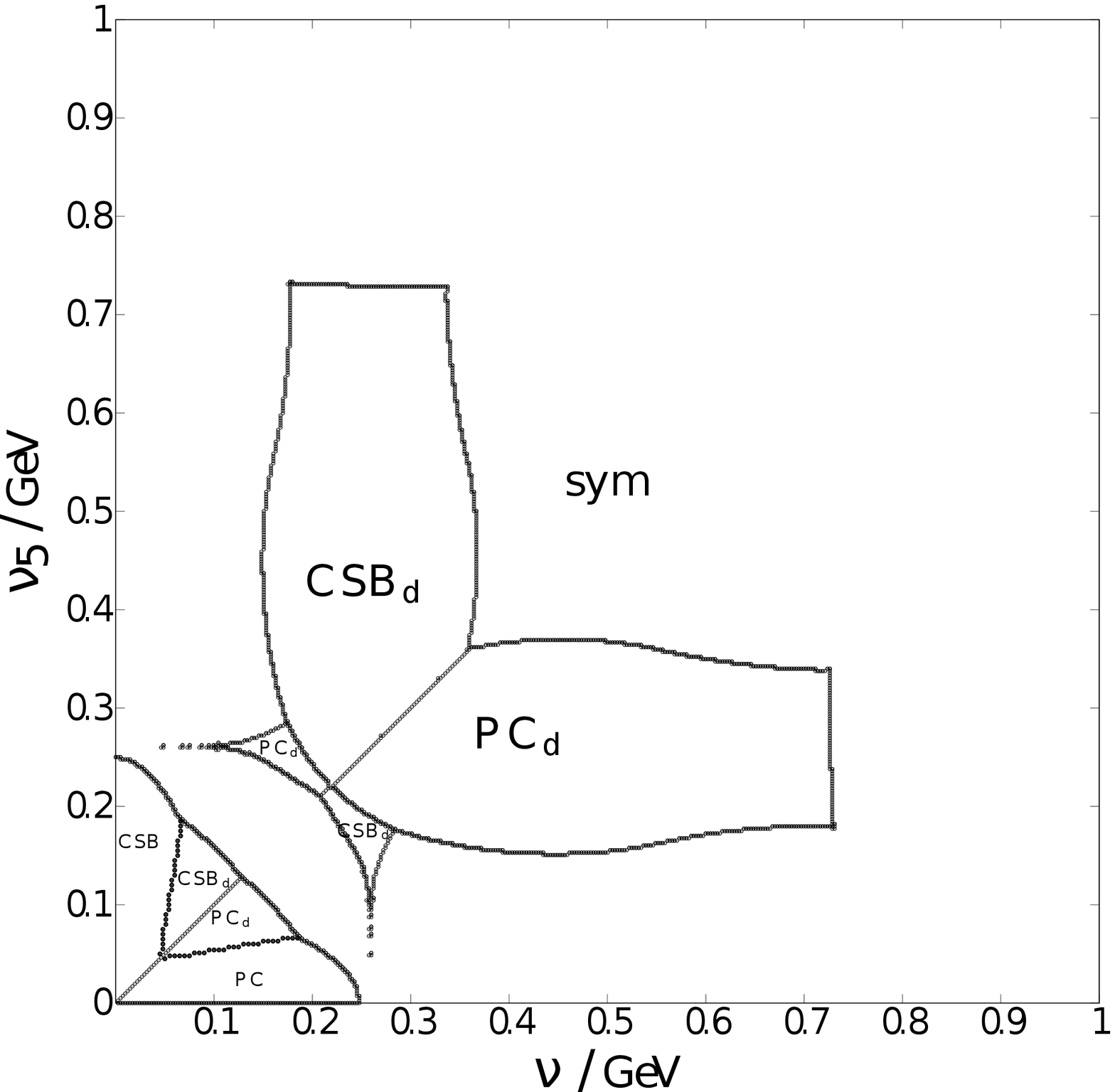}
 \hfill
\includegraphics[width=0.45\textwidth]{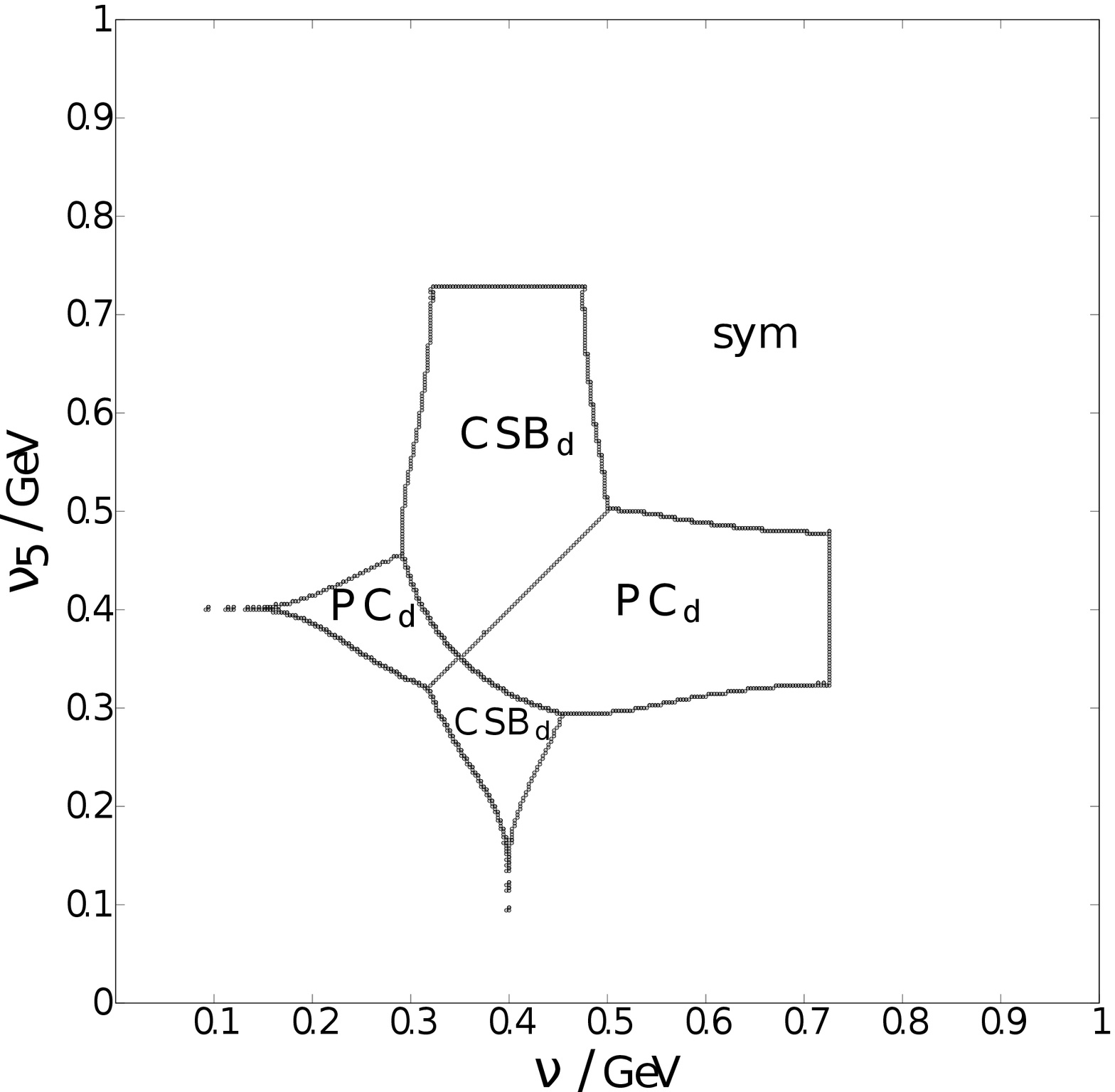}\\
\parbox[t]{0.45\textwidth}{
 \caption{ $(\nu,\nu_{5})$-phase diagram at $\mu=0.26$ GeV.
 All the notations are the same as in Fig. 2.}
 }\hfill
\parbox[t]{0.45\textwidth}{
\caption{ $(\nu,\nu_{5})$-phase diagram at $\mu=0.4$ GeV.  All the notations are the same as in Fig. 2.} }
\end{figure}
\begin{figure}
\includegraphics[width=0.45\textwidth]{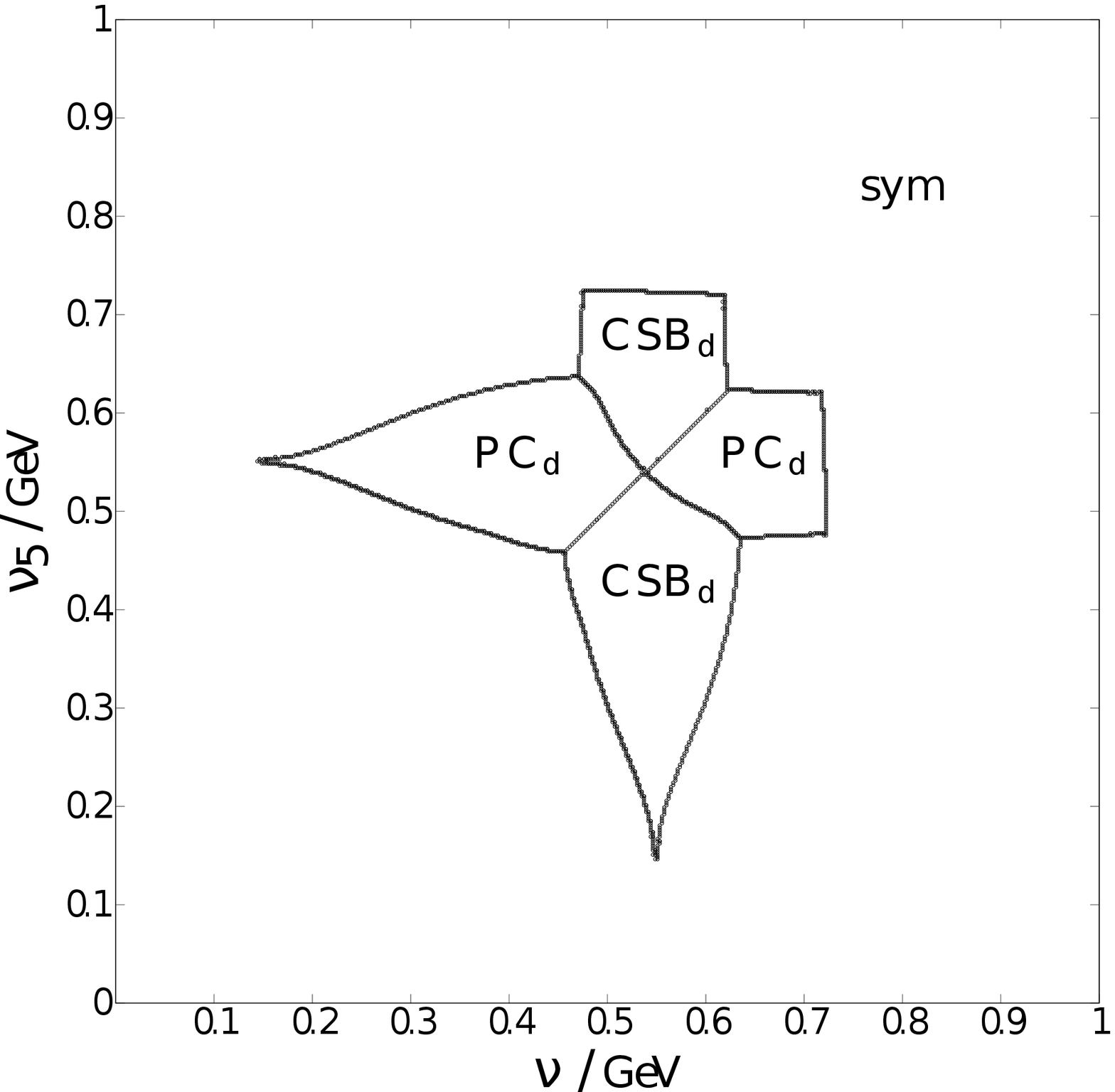}
 \hfill
\includegraphics[width=0.45\textwidth]{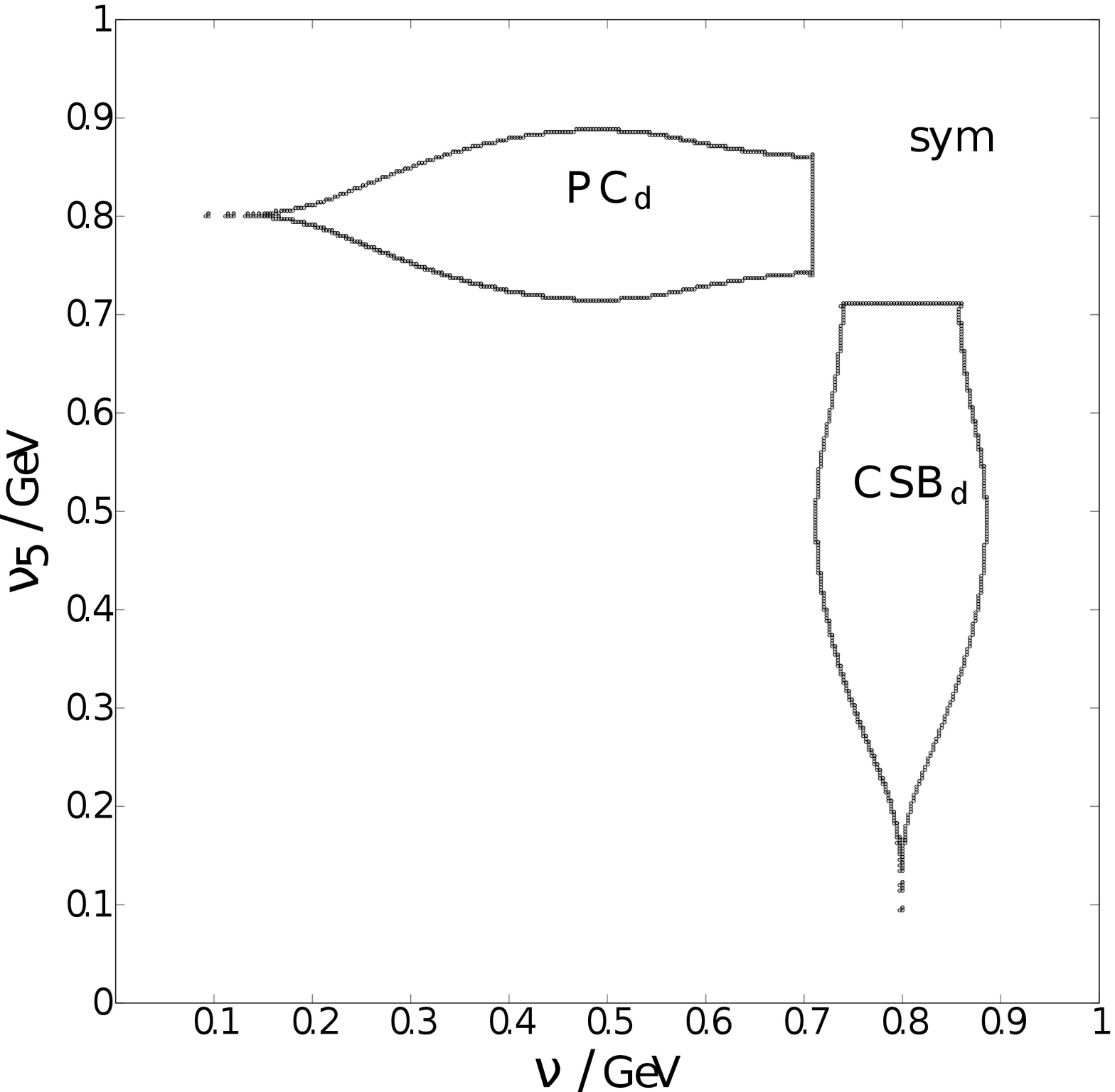}\\
\parbox[t]{0.45\textwidth}{
 \caption{ $(\nu,\nu_{5})$-phase diagram at $\mu=0.55$ GeV.  All the notations are the same as in Fig. 2.}
 }\hfill
\parbox[t]{0.45\textwidth}{
\caption{ $(\nu,\nu_{5})$-phase diagram at $\mu=0.8$ GeV. All the notations are the same as in Fig. 2.} }
\end{figure}

Next, with an increase of $\mu$ (see Figs 4 - 8) the hollow goes into the charged PC phase and PC phase with zero baryon density decreases. It can be noted that charged PC$_{d}$ and CSB$_{d}$ phases of these figures take the form of soles of boots that look towards each other and points at values of $\nu_{5}=\mu$ and $\nu=\mu$, correspondingly. Moreover, with an increase of $\mu$ charged PC$_{d}$ phase and CSB$_{d}$ phase move to the region of larger $\nu_{5}$ and $\nu$, correspondingly. Throughout the movement these two phases go through each other (see Figs 4 -- 7), then separate and run away from each other (Fig. 8).

So one can claim that nonzero $\nu_{5}$ generates the charged pion condensation phase in dense, $n_{B}\ne 0$, quark matter. It is the main result of the paper. Moreover, some phase diagrams of NJL$_4$ model under consideration look qualitatively very similar to the corresponding phase diagrams of NJL$_{2}$ model (compare Fig. 3d in \cite{ekk} and Fig. 8 of the present paper). In (1+1)-dimensional case the charged PC$_d$ phase also goes to higher values of $\nu_{5}$ with increase of $\mu$ but this phase starts at zero values of $\nu$ (see Fig. 3d in \cite{ekk}), whereas in (3+1)-dimensional case the phase starts at some nonzero value of $\nu$ around $0.1$ GeV (see Fig. 8). So in order to realize the charged PC$_{d}$ phase in NJL$_4$ model, besides chiral imbalance there has to be the isotopic imbalance in the system (isospin chemical potential should be nonzero but it does not have to and actually should not be too large either). In that respect one can say that charged PC$_{d}$ phase is generated by both isospin and chiral isospin chemical potentials. This is a new feature that exists only in the NJL$_4$ model, in the NJL$_{2}$ model charged pion condensate phase with nonzero baryon density could be realized just by chiral isospin chemical potential even at $\nu=0$ (see Fig. 2a in Ref. \cite{ekk}).

One can make sure just by looking at the presented
diagrams that (as duality dictates at arbitrary fixed $\mu$) the $(\nu,\nu_5)$-phase diagrams of the model are self-dual, i.e. its CSB and charged PC phases (and CSB$_{d}$ and charged PC$_{d}$ phases) lie mirror-symmetrically with respect to the line $\nu=\nu_5$.

\subsection{$(\nu,\mu)$- and $(\nu_5,\mu)$-phase diagrams}

For a better understanding of the most general $(\mu,\nu,\nu_{5})$-phase diagram of the model, here we consider two other cross-sections of the general diagram, by the planes of $\nu= const$ and $\nu_5= const$.

The $(\nu,\mu)$-phase diagrams of the model at different typical values of $\nu_{5}$ are presented in Figs 9--11. In particular, the phase portrait of the model at zero value of $\nu_{5}$ (see Fig. 9) was first studied in Ref. \cite{eklim}. It has been shown that at $\nu_{5}=0$ there is charged PC phase in the region that spreads along the $\nu$ axis.
But the most of this region is the charged PC phase with zero baryon density. Besides, there is a rather small region of charged PC phase with $n_B\ne 0$ at small $\nu$ and $\mu$ around $0.3$ GeV.

One can consider the phase diagram in Fig. 9 compatible with the
one obtained by lattice QCD technique \cite{Brandt:2016zdy,Son:2000xc}, if take into account the following two remarks.
First, the fact that at rather large values of isotopic chemical potential $\nu$ there is no charged PC phase in Fig. 9 (there is a transition from PC to symmetric phase), which seems to be in contradiction with lattice QCD results, in reality cannot be considered as a contradiction due to the quite obvious observation that NJL$_4$ model is only effective model (low-energy approximation) and can be considered only for chemical potential $\nu$ less than cutoff parameter. Actually, in NJL$_{2}$ model that is renormalizable and can be used for arbitrary high values of $\nu$ there is no transition from the charged PC phase to symmetric one at high isospin imbalance. So this transition could be seen as an artifact of the effective NJL$_4$ model.
Second, charged PC phase in lattice QCD phase diagram starts only from isotopic chemical potential $\nu$ greater than half of pion mass. But our consideration is performed in the chiral limit (zero current quark masses), corresponding to zero pion mass. So, all this makes the phase diagram of Fig. 9 qualitatively the same to the one obtained by lattice QCD simulations (ignoring the fact that on the lattice one cannot go to the zero temperature, as it is in our considerations).

At bigger values of $\nu_{5}$, CSB phase starts to appear from zero values of $\nu$, and the phase transition from it to the charged PC phase takes place at $\nu=\nu_{5}$ (see Fig. 10). As a result, the charged PC$_d$ phase is shifted to greater values of $\nu$ as one increase $\nu_{5}$. Also, there appears a bar of CSB phase that starts from PC phase and goes along the line $\mu=\nu$. This looks very similar to the corresponding phase diagram of the NJL$_2$ model (see Fig. 1a from Ref. \cite{ekk}).
Then, at even larger values of $\nu_{5}$ the bar of CSB phase disappears and charged PC$_d$ phase shifts to the region of larger $\mu$ (see Fig. 11). In this region of $\nu_5$ the shape of the charged PC$_d$ phase resembles again
(as in the case of $(\nu,\nu_{5})$-phase diagram in Fig. 8) a sole of a boot and points towards the value of $\mu=\nu_{5}$.
\begin{figure}
\includegraphics[width=0.45\textwidth]{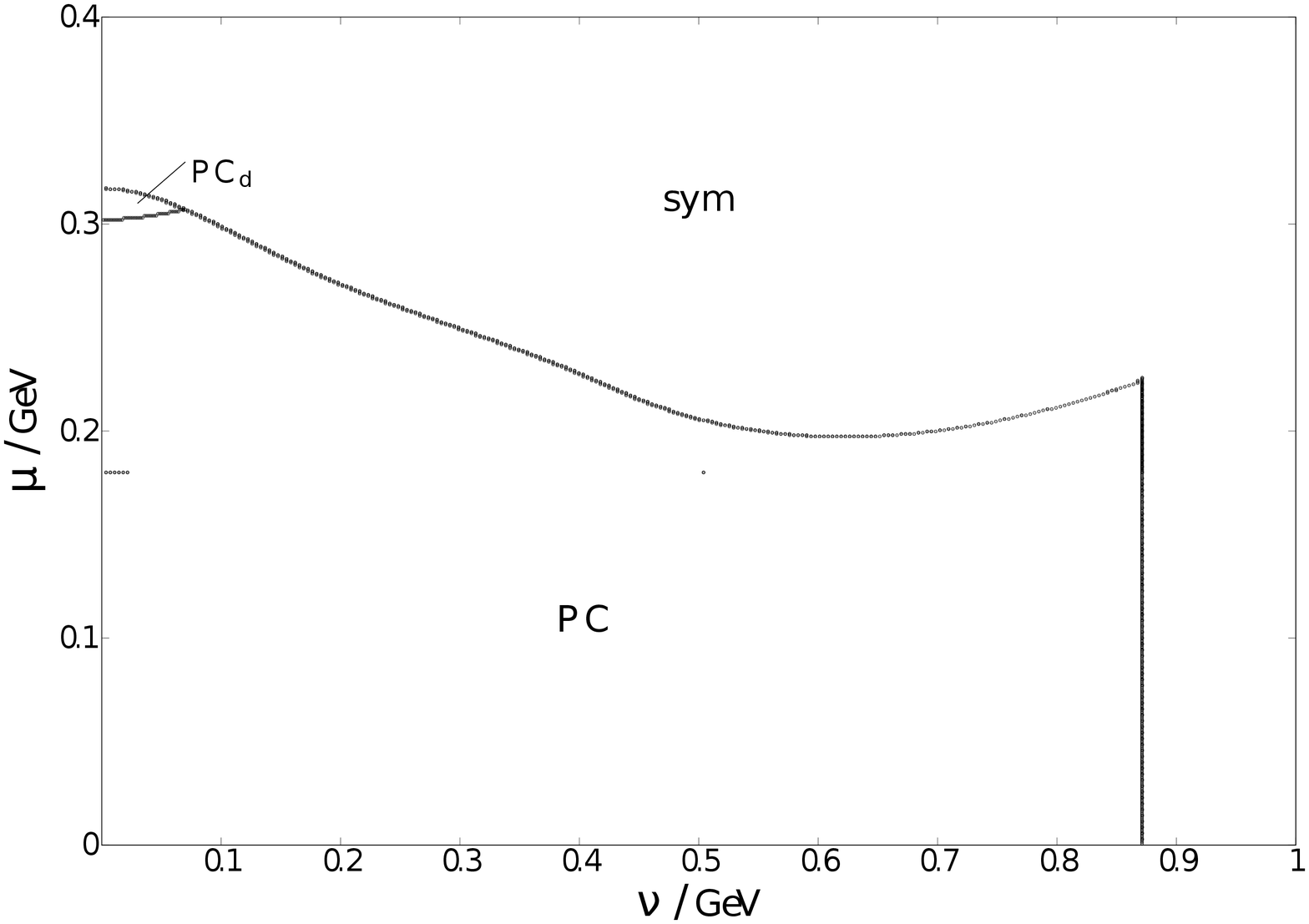}
 \hfill
\includegraphics[width=0.45\textwidth]{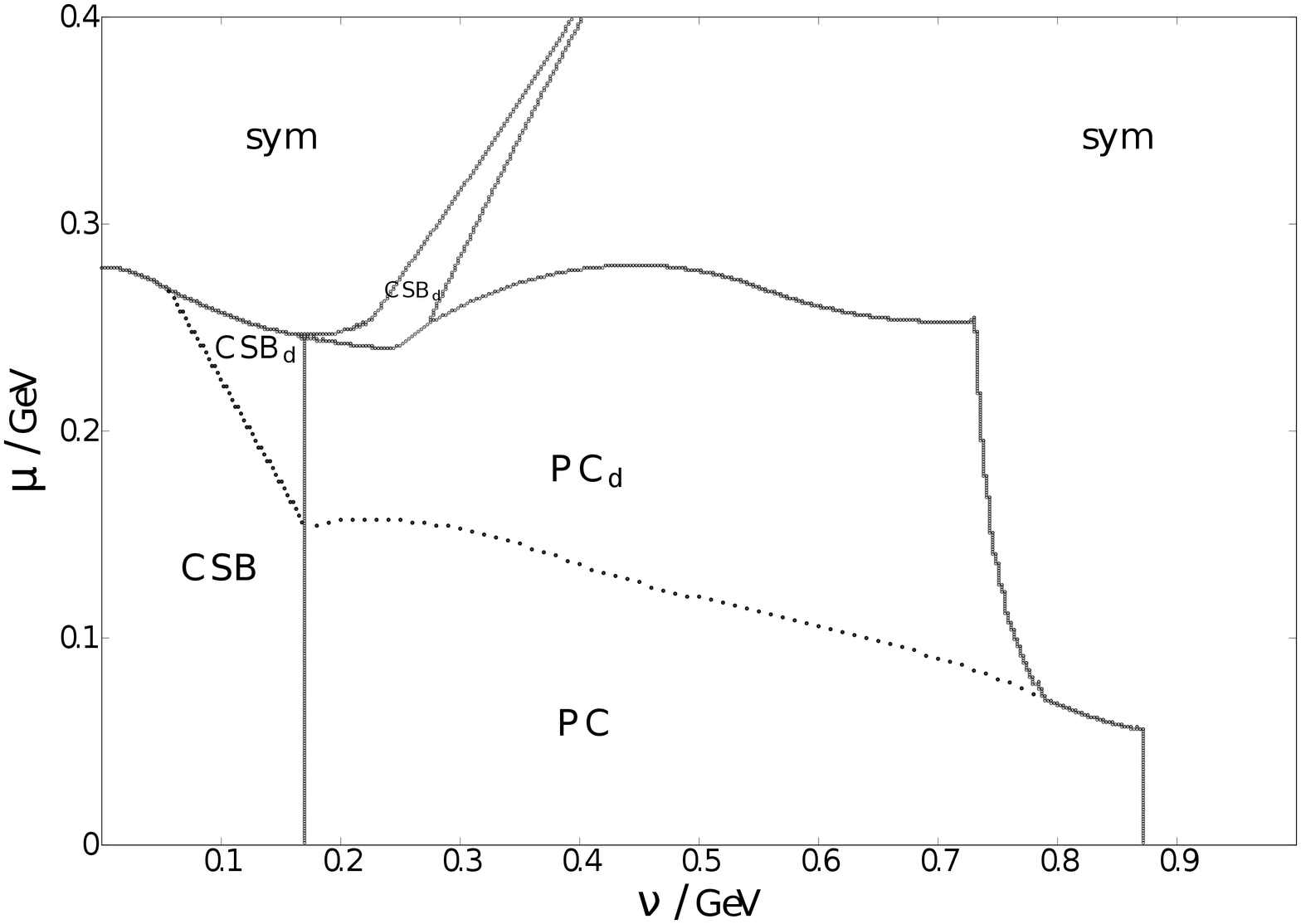}\\
\parbox[t]{0.45\textwidth}{
 \caption{ $(\mu,\nu)$-phase diagram at $\nu_{5}=0$ GeV. All the notations are the same as in Fig. 2.}
 }\hfill
\parbox[t]{0.45\textwidth}{
\caption{  $(\mu,\nu)$-phase diagram at $\nu_{5}=0.17$ GeV. All the notations are the same as in Fig. 2.} }
\end{figure}
\begin{figure}
\includegraphics[width=0.418\textwidth]{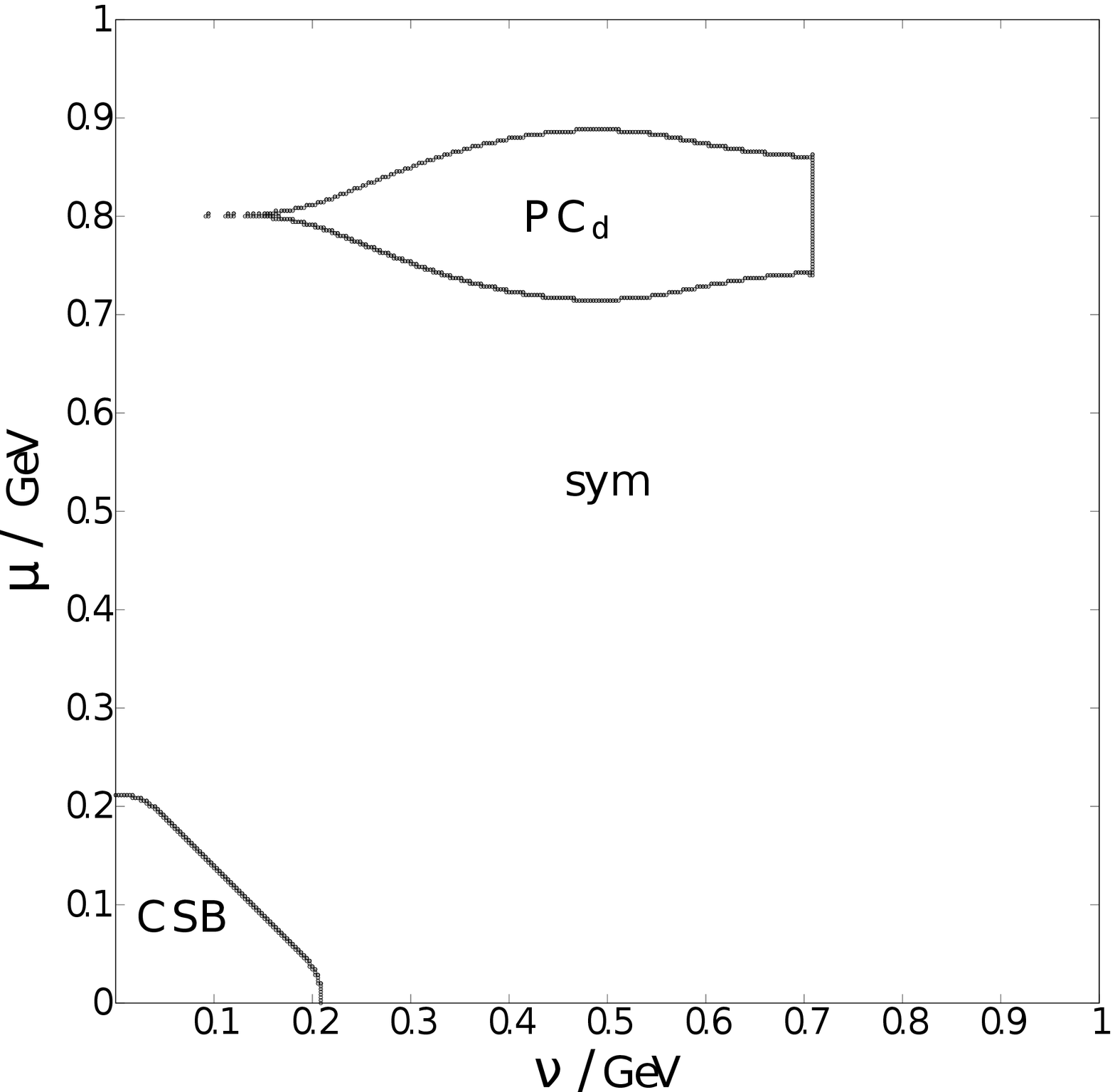}
 \hfill
\includegraphics[width=0.575\textwidth]{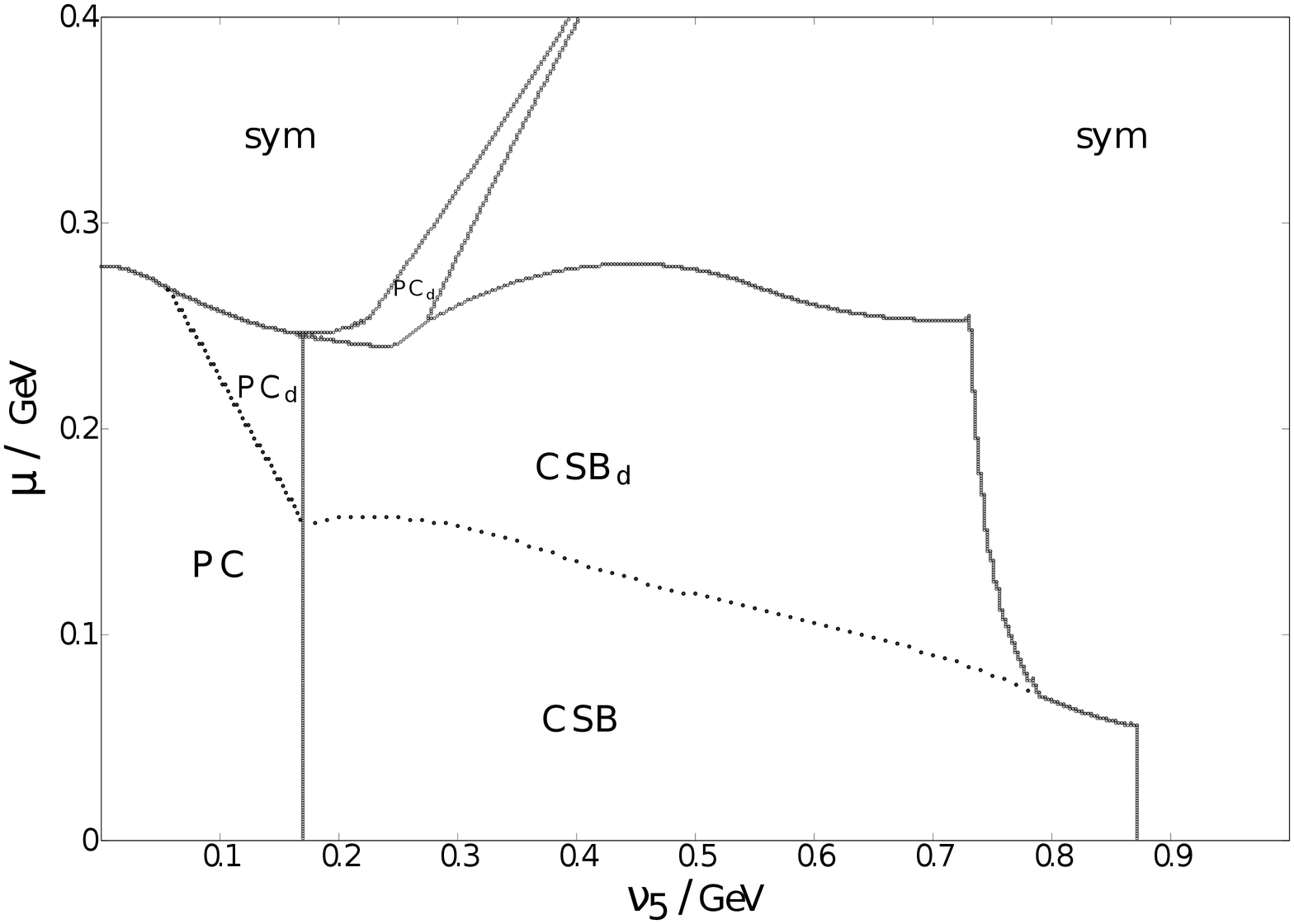}\\
\parbox[t]{0.45\textwidth}{
 \caption{ $(\mu,\nu)$-phase diagram at $\nu_{5}=0.8$ GeV. All the notations are the same as in Fig. 2.}
 }\hfill
\parbox[t]{0.45\textwidth}{
\caption{  $(\mu,\nu_{5})$-phase diagram at $\nu=0.17$ GeV. All the notations are the same as in Fig. 2.} }
\end{figure}

Such an evolution of the $(\nu,\mu)$-phase portraits vs $\nu_5$ (and, in particular, of the charged PC$_d$ phase) in the NJL$_4$ model looks very similar to its corresponding behavior in the NJL$_{2}$ model.
For example see Figs 1a, d in \cite{ekk} and compare them with Figs 10, 11 of the present paper, respectively. The difference is that in NJL$_4$ model the charged PC$_{d}$ phase is realized only for nonzero values of $\nu$, whereas in NJL$_{2}$ model it takes place even for $\nu=0$ (see Fig. 2a in \cite{ekk}). But qualitative behavior is the same. So one can say that in both models chiral imbalance generates charged PC phase in dense ($n_B\ne 0$) quark matter.

Up to now we have presented two types of cross-sections of the general $(\nu,\nu_5,\mu)$-phase diagram, i.e. at $\mu=const$ and $\nu_5=const$. Fortunately, there is no need to perform a detailed numerical calculations in order to find $(\nu_5,\mu)$-phase diagrams of the model at some fixed values of $\nu$. In this case there is a simpler way, which is based on the duality invariance (\ref{16}) of the TDP. So one can apply the duality transformation $\cal D$ (\ref{16}) to the $(\nu,\mu)$-phase diagram at arbitrary fixed $\nu_5=A$, in order to obtain the dually conjugated phase diagram, which is nothing more than a $(\nu_5,\mu)$-phase diagram at fixed $\nu=A$. Hence, to find, e.g., the $(\nu_5,\mu)$-phase diagram at $\nu=0.17$ GeV, we should start from the corresponding $(\nu,\mu)$-phase diagram at fixed $\nu_5=0.17$ GeV of Fig. 10 and make the simplest replacement of the notations in this figure: $\nu\leftrightarrow\nu_5$, PC$_{d}\leftrightarrow$CSB$_{d}$ and PC$\leftrightarrow$CSB (note, the symmetric phase is intact under the dual transformation). The result is the $(\nu_5,\mu)$-phase portrait of the model at $\nu=0.17$ GeV (see Fig. 12). \vspace{0.4cm}

\subsection{Qualitative arguments in favor of the charged PC$_d$ phase}

In the present subsection we would like to demonstrate, using some qualitative condensed matter physics arguments, the principle possibility of the charged pion condensation phenomenon in dense quark medium. The system is characterized by several chemical potentials, including the chiral chemical potential responsible for the chiral imbalance of quark matter. In this case there are different densities for the left- and right-handed quarks. So, in the energy-momentum space left- and right-handed particles occupy unequal regions (Fermi seas). As a result, the Fermi surfaces (which can be identified with corresponding particle number chemical potentials) are different for left- and right-handed quarks. In general, we have studied the phase structure of the initial NJL model (1) in terms of $\mu,\nu,\nu_{5}$ chemical potentials. However, sometimes it is convenient to perform the consideration in terms of the quantities $\mu_{uR},\mu_{uL},\mu_{dR},\mu_{dL}$ which are the chemical potentials for right- and left-handed $u$ and $d$ quarks, respectively. To find these chemical potentials, we should introduce the left- and right-handed $u,d$-quark fields,
\begin{eqnarray}
q_{uR}=\frac{1+\gamma^5}{2}q_u,~~~ q_{uL}=\frac{1-\gamma^5}{2}q_u,~~~q_{dR}=\frac{1+\gamma^5}{2}q_d,~~~ q_{dL}=\frac{1-\gamma^5}{2}q_d.\label{B1}
\end{eqnarray}
Then, using Eq. (\ref{B1}), the chemical potential term of Eqs (\ref{1}) and (\ref{2}) can be presented in the following form
\begin{eqnarray}
\bar q\big [\mu\gamma^0
+ \nu\tau_3\gamma^0+\nu_{5}\tau_3\gamma^0\gamma^5\big ]q&=&\bar q_{uR}\gamma^0q_{uR}\big (\mu+\nu+\nu_5\big )+\bar q_{uL}\gamma^0q_{uL}\big (\mu+\nu-\nu_5\big )\nonumber\\
&+&\bar q_{dR}\gamma^0q_{dR}\big (\mu-\nu-\nu_5\big )+\bar q_{dL}\gamma^0q_{dL}\big (\mu-\nu+\nu_5\big ).
\label{B2}
\end{eqnarray}
It is clear from Eq. (\ref{B2}) that particle number chemical potentials of
the left- and right-handed $u,d$-quark fields $q_{uR},q_{uL},q_{dR},q_{dL}$ are the following,
\begin{eqnarray}
\mu_{uR}=\mu+\nu+\nu_5,~~\mu_{uL}=\mu+\nu-\nu_5,~~\mu_{dR}=
\mu-\nu-\nu_5,~~\mu_{dL}=\mu-\nu+\nu_5,
\label{B3}
\end{eqnarray}
respectively. In addition, one can present the auxiliary scalar fields of Eq. (\ref{200}) in terms of left- and right-handed $u,d$-quark fields,
\begin{eqnarray}
\bar qq\sim \bar q_{uR}q_{uL}+\bar q_{uL}q_{uR}+(u\leftrightarrow d),~~
\pi^+\sim\bar q_{dL}q_{uR}-\bar q_{dR}q_{uL},...
\label{B4}
\end{eqnarray}
Here we are going to present two  qualitative quark-antiquark pairing mechanisms, leading to the isospin $U_{I_3}(1)$ symmetry breaking and to the charged PC phenomenon in dense quark matter described by massless Lagrangian (1). Our consideration is a generalization to the case of several chemical potentials of the approach of the paper \cite{kojo}, where a qualitative analysis of possible condensed phenomena in a dense quark medium was carried out. For simplicity, we suppose that isospin asymmetry of dense ($\mu\ne 0$) weakly interacting quark matter is very small, $\nu\ll \mu,\nu_5$ and, in addition, $\mu\gtrsim\nu_5$.

In the first mechanism of isospin symmetry breaking only the excitations around the Fermi surfaces (chemical potentials) of the $q_{uL}$ and $q_{dR}$ quarks are involved. By assumption, we see that $\mu_{uL}\approx \mu_{dR}\approx\mu-\nu_5\approx 0$ (see Eq. (\ref{B3})). (But the chemical potentials of the $q_{uR}$ and $q_{dL}$ quarks, i.e. $\mu_{uR}\approx \mu_{dL}\approx\mu+\nu_5$, are rather large in this case.) Then under the influence of external weak effects, the $q_{uL}$ and $q_{dR}$ quarks with momentum $\vec p$ can appear just above the Fermi surface, which corresponds to the Fermi energy $\epsilon_F=\mu-\nu_5$. In this case below this Fermi surface two holes appear, which are the antiparticles $\bar q_{uL}$ and $\bar q_{dR}$ with momenta $-\vec p$. \footnote{Since in our consideration all $q_{uL}$- and $q_{dR}$-quark excitations are massless and their energies are near the Fermi surface $\epsilon_{F}$ by assumption, we have $|\vec p|=\epsilon_{F}=\mu-\nu_5\approx 0$ for all such $q_{uL}$ and $q_{dR}$ excitations.} The relative momentum between the quark $q_{uL}$ and the hole $\bar q_{dR}$ is a rather small quantity (since $|\vec p|\approx 0$), so they can easily form, due to the weak attractive interaction in the particle-hole pair having the same quantum numbers with those of pions, $\bar{q}_{dR} q_{uL} \sim \pi^+$ (see Eq. (39)), with zero net momentum. The condensation of these $\pi^+$ pairs rearranges initial ground state of the system in favor of the new ground state in which both isospin $U_{I_3}(1)$ symmetry (see Eq. (2)) and spatial parity $\cal P$ are spontaneously broken. Moreover, this condensate is a spatially {\it homogeneous} one. Since the quark number chemical potential $\mu$ is a rather large quantity corresponding to nonzero quark number density, we obtain in this case the realization of the charged PC phase in dense quark matter.

Similar arguments can help us to establish that a condensation of the $\bar qq$ pairs is forbidden in this case (i.e. at $\nu\ll \mu,\nu_5$ and $\mu\gtrsim\nu_5$) by kinematic reason. Indeed, the right-handed $q_{uR}$ quark can also be born just above its Fermi surface $\mu_{uR}\approx \mu+\nu_5$ with some momentum $\vec p^{~\prime}$ such that $|\vec p^{~\prime}|\approx\mu+\nu_5$. However, the relative momentum $\vec p^{~\prime}+\vec p$ between this $q_{uR}$ quark and the corresponding left-handed $\bar q_{uL}$-quark excitation with momentum $-\vec p$ has a rather large absolute value, $|\vec p^{~\prime}+\vec p|\sim \mu_{uR}-\mu_{uL}\approx 2\mu_5\sim\mu$. So the probability for creation of the  $\bar q q\sim \bar q_{uL}q_{uR}$ pair (see Eq. (\ref{B4})) is very small (due to a large relative momentum, quarks will not have enough time for pairing). Hence in this case the appearance of the chiral condensate is suppressed.

The above qualitative arguments in favor of the charged PC$_d$ phase are well illustrated by the phase portrait of Fig. 12, where in the region $\{\nu\ll \mu,\nu_5;~\mu\approx\nu_5\}$ just the PC$_d$ phase is arranged. In addition, the arguments are quite suited for a qualitative explanation of the phase diagrams Figs 1d and 2a of the Ref. \cite{ekk} in the case of the NJL$_2$ model, as well.

Since the model under consideration is invariant under the duality transformation (21), we can use the same qualitative arguments in order to explain why in the dually conjugated region $\{\nu_5\ll \mu,\nu;~\mu\approx\nu\}$ the CSB$_d$ phase is realized (see Fig. 10).

Let us now discuss the second way (mechanism) of the isospin symmetry breaking, in which already the excitations around the Fermi surfaces of the $q_{uR}$ and $q_{dL}$ quarks play a decisive role. As before, we suppose that $\nu\ll \mu,\nu_5$ and $\mu\gtrsim\nu_5$, so the Fermi surfaces (chemical potentials) of these quarks are approximately equal, $\mu_{uR}\approx \mu_{dL}\approx\mu+\nu_5$ (see Eq. (\ref{B3})). Then arbitrary small external excitements can cause the appearance of the $q_{uR}$ quark with momentum $\vec p$ and the $q_{dL}$ quark with momentum $-\vec p$ just above the Fermi surface $\varepsilon_F=\mu+\nu_5$. In this case below the Fermi surface $\varepsilon_F$ two holes appear, which are the antiparticles $\bar q_{uR}$ and $\bar q_{dL}$ with momenta $-\vec p$ and $\vec p$, respectively. Since the relative momentum between the $q_{uR}$ quark and $\bar q_{dL}$ antiquark is zero, they can easily form, due to a weak attractive interaction in the quark-antiquark channel, the $\pi^+\sim\bar q_{dL}q_{uR}$ pair (see Eq. (\ref{B4})). The condensation of these $\pi^+$ pairs also results in the
spontaneous isospin $U_{I_3}(1)$ symmetry breaking and appearing of the PC phase in dense quark matter. However, because of the nonzero net momentum $2\vec p$ of this $\pi^+\sim\bar q_{dL}q_{uR}$ pair, where $|\vec p|\sim\mu+\mu_5$, the $\pi^+$ condensation is a spatially {\it inhomogeneous} in this case (for details see Ref. \cite{kojo}).

Recall, in the present paper we investigate the phase structure of the initial NJL model, supposing that all condensates are spatially homogeneous, i.e. we ignore the second (inhomogeneous) way for quark pairing. However, in future we are going to study the competition of these mechanisms and take into account the possibility of inhomogeneous both CSB and PC condensates in the framework of this model. \vspace{0.4cm}

At the end of this section we would like, based on the latest both theoretical and experimental studies, to discuss in more details the modern status of the charged PC$_d$ phase. Earlier, in the Introduction we have already noticed that this phase is not predicted in electrically neutral dense quark matter within the framework of ordinary NJL$_4$ model (with nonzero bare quark mass) without chiral imbalance \cite{andersen}. This conclusion is also supported by Ref. \cite{ohnishi}, where it was shown that $s$-wave (i.e. homogeneous) charged pion condensation is suppressed in neutron stars with hyperons. \footnote{The study of the in-medium pion properties in this paper is based, in particular, on experimental observations of deeply bound atomic states of $\pi$ mesons in some isotopes  \cite{suzuki}.} On the other hand, if an external magnetic field as well as the rotation of a dense medium are taken into account, then the charged PC phase can be observed both in neutron stars and in heavy-ion collisions \cite{liu}. Our present results are not directly applicable to neutron stars (since in our consideration the condition of the electric neutrality of the medium is absent and quarks have zero bare mass), but predict the possibility of the parity $\cal P$- or ${\cal CP}$-breaking phase (in our case it is the PC$_d$ phase) of the quark-matter fireball in heavy-ion collision. The similar conclusion is obtained in some recent papers (see, e.g., in Ref. \cite{kaw}), when chiral imbalance of quark matter is taken into account.

\section{Summary and Conclusions}

In this paper the influence of isotopic and chiral imbalance on phase structure of dense quark matter has been investigated in the framework of the (3+1)-dimensional NJL model with two quark flavors in the large-$N_{c}$ limit ($N_{c}$ is the number of colors). Dense matter means that our consideration has been performed at nonzero baryon $\mu_B$ chemical potential. Isotopic and chiral imbalance in the system were accounted for by introducing isospin $\mu_I$ and chiral isospin $\mu_{I5}$ chemical potentials (see Lagrangian (1)). All the results are obtained in the chiral limit.

Earlier, analogous phase structure was considered in the framework of massless NJL$_{2}$ model in Refs \cite{ekk,kkzz}, where it was shown that $\mu_{I5}$ promotes charged PC phase with {\it nonzero baryon density} (in Refs \cite{ekk,kkzz} and in the present consideration this phase is denoted as charged PC$_d$ phase). Although the NJL$_{2}$ model captures the main features of QCD and can be used as a toy model for the qualitative description of specific properties of QCD, it is not in any sense guaranteed that it will succeed in every case. So we study the more realistic QCD effective field theory in order to put our NJL$_2$ results on a more solid basis. It appears that the phase diagrams of the two models are very similar and the main result that chiral isospin chemical potential generates charged PC phenomenon in dense quark matter holds both in the NJL$_{2}$ and NJL$_4$ models.

Studies in the framework of NJL$_{2}$ and more realistic (3+1)-dimensional NJL model in a sense complement each other. For example, NJL$_4$ model predicts that at rather large isotopic chemical potential (actually outside of the scope of validity of NJL$_4$ model) there is a phase transition back to a symmetric phase. But we know that there is no such transition in the lattice simulations of QCD. In contrast, the NJL$_{2}$ model does not predict such a transition. So in the region, where NJL$_4$ is not a reliable theory and gives wrong results, one can use predictions of the NJL$_{2}$ model. Or one can see that in the NJL$_4$ model at high values of baryon chemical potential $\mu_B$ the charged PC$_d$ phase shifts up to higher values of chiral isospin chemical potential $\mu_I$ and at some $\mu_B$ it goes outside of a region of validity of NJL$_4$ model. But a similar result has been obtained in the NJL$_{2}$ model as well, where there are no constraints on chemical potential values, i.e. one can consider the NJL$_{2}$ model at arbitrary high values of chemical potentials. So one can think that it is possible to go beyond the scope of validity of the NJL$_4$ model and trust the result in this case. However, as it has been said earlier, the NJL$_4$ model is a more realistic theory for QCD and in the region of validity its results are more trustworthy and corroborate the ones of the NJL$_{2}$ model. Furthermore, in the case of zero baryon chemical potential our results resemble the ones obtained in the lattice simulations for real QCD. Namely, chiral imbalance promotes the chiral symmetry breaking phenomenon.

Let us summarize the most essential results of our paper obtained in the chiral limit.

1) Chiral isospin
chemical potential generates charged pion condensation in dense quark matter in the framework of (3+1)-dimensional NJL model. So this phenomenon is predicted in two models, in NJL$_4$ and NJL$_{2}$, and might be the property of real QCD.

2) It has been also demonstrated that in the framework of the NJL$_4$
model duality correspondence between CSB and charged PC phenomena
takes place in the leading order of the large-$N_c$ approximation as in NJL$_{2}$ model.

3) In contrast to NJL$_{2}$ model, where the generation of the PC$_{d}$ phase occurs even at very small values of isospin chemical potential $\nu$, the generation of the PC$_{d}$ phase in NJL$_4$ model requires not very large but nonzero isospin chemical potential. So charged pion condensation phenomenon cannot occur in the absence of isotopic imbalance in the system and in order to generate PC$_{d}$ phase one needs to have both nonzero isospin $\mu_I$ and chiral isospin $\mu_{I5}$ chemical potentials.

As we discussed in the Introduction the dualities akin to ours were obtained in the framework of universality principle (large $N_{c}$ orbifold equivalence) of phase diagrams in QCD and QCD-like theories in the limit of large $N_{c}$ \cite{Cherman,1103.5480,Hanada:2011ju,Kashiwa:2017yvy,han}.
So several methods (NJL model considerations and orbifold equivalence) point to the similar dualities of the phase portrait of QCD. Are there such dualities in the lattice QCD?
It has been mentioned earlier that introducing isotopic chemical potential or chiral chemical potential does not lead to the sign problem. And we believe that our results can be supported by lattice QCD investigations at least in the case of a zero baryon chemical potential (and nonzero isotopic or chiral isotopic chemical potential). Moreover, we hope that our results might shed  new light on phase structure of dense quark matter with isotopic and chiral imbalance and hence could be of importance for describing physics in the heavy ion collision experiments. Since dense quark matter with isotopic and chiral imbalance can be created in the fireball after a collision of heavy nuclei.

\appendix{}

\section{Calculation of roots of $P_{\pm}(\eta)$}

In this appendix it will be shown how to get roots of the following quartic equation (general quartic equation could be reduced to the one of this form)
\begin{eqnarray}
P_+(\eta)\equiv \eta^4-2a\eta^2+b\eta+c=0.\label{A1}
\end{eqnarray}
The coefficients $a,b,c$ in Eq. (\ref{A1}) are given by the relations (\ref{10}). First, we represent the polynomial on the left-hand side of this equation as the product of two quadratic polynomials,
\begin{eqnarray}
(\eta^2+r\eta+q)(\eta^2-r\eta+s)=0,\label{A2}
\end{eqnarray}
where
$$
-r^{2}+q+s=-2a,\,\,\,\,\,\,\,\,\,\,\,\,\,qs=c,\,\,\,\,\,\,\,\,\,\,\,\,\,\,\,rs-rq=b.
$$
It follows from these relations that
\begin{eqnarray}
q=\frac{1}{2} \left(-2 a+r^2-\frac{b}{r}\right),~~~s=\frac{1}{2} \left(-2 a+r^2+\frac{b}{r}\right).\label{A3}
\end{eqnarray}
Substituting Eq. (\ref{A3}) into Eq. (\ref{A2}), one gets
that $r=\sqrt{R}$, where $R$ is one of the solutions of the following cubic equation
\begin{equation}
X^{3}+AX=BX^{2}+C,
\label{cub13}
\end{equation}
where we used notations $A, B, C$ that are given by
$$
A=16(\Delta ^2 {\nu _5}^2+ \nu ^2{ \nu _5}^2+ \nu ^2 M^2+ p^2 \left(\nu ^2+{\nu _5}^2\right)),~~~B=4a,~~~C=b^2.
$$
All three solutions of the cubic equation (\ref{cub13}) are
\begin{equation}
R_{1,2,3}=\frac{1}{3} \left(4 a+\frac{L}{\sqrt[3]{J}}+\sqrt[3]{J}\right),\label{A5}
\end{equation}
where
$$
J=\frac{1}{2}(K+i\sqrt{4 L^3-K^2}),~~K=128 a^3-36 a A+27 b^2,~~L=-3A+16 a^2,
$$
and $ \sqrt[3]{J}$ in Eq. (\ref{A5}) means each of three possible complex valued roots. There is a determinant $D\equiv 4L^{3}-K^{2}>0$ of the equation (\ref{cub13}) that can tell us the structure of roots $R_{1,2,3}$. Namely,
if $D>0$ then all roots $R_i$ are real and different, if $D=0$ all roots are real and at least two are equal. Finally, if $D<0$ then one root is real and two are complex conjugate. So, there is always a real solution of Eq. (\ref{cub13}). In numerical simulations it is more handy to work with real solution and it is always possible to choose one.
There is a procedure that, depending on values of parameters, chooses a real solution, but it is quite lengthy so we will not present it here.

And when one has found $r$, the roots of Eq. (\ref{A1}) has the following form
\begin{equation}
\eta_{1}=\frac{1}{2} \left(-\sqrt{r^2-4 q}-r\right),
\eta_{2}=\frac{1}{2} \left(\sqrt{r^2-4 q}-r\right),
\eta_{3}=\frac{1}{2} \left(r-\sqrt{r^2-4 s}\right),
\eta_{4}=\frac{1}{2} \left(r+\sqrt{r^2-4 s}\right). \label{A6}
\end{equation}
The roots $\eta_{5,6,7,8}$ of the equation $P_-\equiv\eta^4-2a\eta^2-b\eta+c=0$ can be obtained by changing $b\to-b$ in Eq. (\ref{A1}) (or $q\leftrightarrow s$ in Eq. (\ref{A6}) with $r$ unchanged). So, we have
 $$
\eta_{5}=-\eta_{4},\,\eta_{6}=-\eta_{3},\,\eta_{7}=-\eta_{2},\,
\eta_{8}=-\eta_{1}.
$$

\end{document}